\documentclass[journal]{IEEEtran}

\usepackage{graphicx}
\usepackage{balance}
\usepackage{url}
\usepackage[acronym,toc,shortcuts]{glossaries}
\usepackage{epsfig}
\usepackage{amssymb}
\usepackage{amsthm}
\usepackage{multirow}
\usepackage{caption}
\usepackage{subcaption}
\usepackage{siunitx}
\usepackage{soul}
\usepackage{algorithm}
\usepackage{algpseudocode}
\usepackage{csquotes}
\usepackage{bbm}
\usepackage{dsfont}
\usepackage{xcolor}
\usepackage{enumitem}
\usepackage{adjustbox}
\usepackage{array}
\usepackage{multirow}
\usepackage{fancyhdr}
\usepackage[T1]{fontenc}

\usepackage{tabularx,ragged2e,booktabs}
\newcolumntype{L}{>{\RaggedRight\arraybackslash}X}

\makeatletter

\newacronym{2D}{2D}{two dimensional}
\newacronym{3D}{3D}{three dimensional}
\newacronym{5G}{5G}{Fifth Generation}
\newacronym{5G PPP}{5G PPP}{5G Infrastructure Public Private Partnership}
\newacronym{3GPP}{3GPP}{3rd Generation Partnership Project}
\newacronym{4D}{4D}{four dimensional}
\newacronym{AAA}{AAA}{Authentication, Authorization and Accounting}
\newacronym{ABSF}{ABSF}{Almost-Blank Subframe}
\newacronym{AES}{AES}{Advanced Encryption Standard }
\newacronym{AMC}{AMC}{Adaptive Modulation and Coding}
\newacronym{AP}{AP}{access point}
\newacronym{API}{API}{Application Programming Interface}
\newacronym{APN}{APN}{Access Point Name}
\newacronym{AWGN}{AWGN}{additive white Gaussian noise}
\newacronym{BBU}{BBU}{Baseband Unit}
\newacronym{BE}{BE}{best-effort}
\newacronym{BET}{BET}{Blind Equal Throughput}
\newacronym{BLAST}{BLAST}{Bell Laboratories Layered Space-Time}
\newacronym{BLER}{BLER}{Block Error Rate}
\newacronym{BS}{BS}{base station}
\newacronym{BTP}{BTP}{Backhaul Transport Provider}
\newacronym{BTS}{BTS}{Base Transceiver Station}
\newacronym{CA}{CA}{carrier aggregation}
\newacronym{MBAR}{MBAR}{Mobile Backhaul Aggregation Router}
\newacronym{CAPEX}{CapEx}{capital expenditure}
\newacronym{CDF}{CDF}{Cumulative Distribution Function}
\newacronym{CELL-ID}{CELL-ID}{cell identification ID}
\newacronym{CIO}{CIO}{cell individual offset}
\newacronym{CDN}{CDN}{Content Delivery Network}
\newacronym{CN}{CN}{core network}
\newacronym{CP}{CP}{cyclic prefix}
\newacronym{CPU}{CPU}{central processing unit}
\newacronym{CoMP}{CoMP}{Coordinated Multipoint}
\newacronym{CSR}{CSR}{Cell Site Router}
\newacronym{CQI}{CQI}{Channel Quality Indicator}
\newacronym{C-RAN}{C-RAN}{Cloud RAN}
\newacronym{CS}{CS}{central scheduler}
\newacronym{CSI}{CSI}{channel state information}
\newacronym{CRE}{CRE}{cell range expansion}
\newacronym{D2D}{D2D}{Device-to-Device}
\newacronym{DL}{DL}{Downlink}
\newacronym{DFT}{DFT}{discrete Fourier transform}
\newacronym{DSL}{DSL}{Digital subscriber line}
\newacronym{EARFCN}{EARFCN}{E-UTRA Absolute Radio Frequency Channel Number}
\newacronym{EC}{EC}{European Commission}
\newacronym{eICIC}{eICIC}{enhanced inter-cell interference cancellation}
\newacronym{eMBB}{eMBB}{enhanced Mobile Broadband}
\newacronym{eNodeB}{eNodeB}{Evolved Node B}
\newacronym{EPC}{EPC}{Evolved Packet Core}
\newacronym{EPS}{EPS}{Evolved Packet System}
\newacronym{ETSI}{ETSI}{European Telecommunications Standards Institute}
\newacronym{E-UTRAN}{E-UTRAN}{Evolved Universal Terrestrial Radio Access Network}
\newacronym{FDMA}{FDMA}{frequency division multiple access}
\newacronym{FFT}{FFT}{fast Fourier transform}
\newacronym{FTP}{FTP}{File Transfer Protocol}
\newacronym{FU}{FU}{Frame Usage}
\newacronym{GTP}{GTP}{GPRS Tunneling Protocol}
\newacronym{GGSN}{GGSN}{Gateway GPRS Support Node}
\newacronym{GPS}{GPS}{global positioning system}
\newacronym{GRA}{GRA}{Grey relational analysis}
\newacronym{GSM}{GSM}{Global System for Mobile Communications}
\newacronym{GEO}{GEO}{Geosynchronous}
\newacronym{GTP-U}{GTP-U}{GPRS Tunneling Protocol-User Plane}
\newacronym{HDFS}{HDFS}{Hadoop Distributed File System}
\newacronym{HetNet}{HetNet}{Heterogeneous Network}
\newacronym{HiveQL}{HiveQL}{Hive Query language}
\newacronym{HD}{HD}{High Definition}
\newacronym{HEO}{HEO}{High Earth Orbit}
\newacronym{HO}{HO}{handover}
\newacronym{HARQ}{HARQ}{Hybrid automatic repeat request}
\newacronym{HS-DSCH}{HS-DSCH}{High Speed Downlink Shared Channel}
\newacronym{HSS}{HSS}{Home Subscriber Station}
\newacronym{HTS}{HTS}{High Throughput Satellite}
\newacronym{HTTP}{HTTP}{Hypertext Transfer Protocol}
\newacronym{ICIC}{ICIC}{inter-cell interference cancellation}
\newacronym{ICN}{ICN}{information-centric network}
\newacronym{IEEE}{IEEE}{Institute of Electrical and Electronics Engineers}
\newacronym{IMEI}{IMEI}{International Mobile Station Equipment Identity}
\newacronym{IMSI}{IMSI}{International Mobile Subscriber Identity}
\newacronym{IMS}{IMS}{IP Multimedia Subsystem}
\newacronym{IMT-A}{IMT-A}{International Mobile Telecommunications - Advanced}
\newacronym{ITU}{ITU}{International Telecommunication Union}
\newacronym{IP}{IP}{Internet Protocol}
\newacronym{IPsec}{IPsec}{Internet Protocol Security}
\newacronym{IoT}{IoT}{Internet of Things}
\newacronym{JSON}{JSON}{JavaScript Object Notation}
\newacronym{KPI}{KPI}{key performance indicator}
\newacronym{LAC}{LAC}{location area code}
\newacronym{LEO}{LEO}{Low Earth Orbit}
\newacronym{LOS}{LOS}{Line-of-Sight}
\newacronym{LTE}{LTE}{Long Term Evolution}
\newacronym{LTE-A}{LTE-A}{Long Term Evolution Advanced}
\newacronym{mmWave}{mmWave}{millimeter wave}
\newacronym{MAC}{MAC}{Medium Access Control}
\newacronym{MADM}{MADM}{Multiple Attribute Decision Making}
\newacronym{MBH}{MBH}{Mobile Backhaul}
\newacronym{MCS}{MCS}{Modulation Coding Scheme}
\newacronym{MEO}{MEO}{Medium Earth Orbit}
\newacronym{MEW}{MEW}{multiplicative exponent weighting}
\newacronym{MIMO}{MIMO}{multiple-input multiple-output}
\newacronym{MME}{MME}{Mobility Management Entity}
\newacronym{mMTC}{mMTC}{massive Machine Type Communications}
\newacronym{MMF}{MMF}{max-min fairness}
\newacronym{MMSE}{MMSE}{minimum mean square error}
\newacronym{MPLS}{MPLS}{Multiprotocol Label Switching}
\newacronym{MSISDN}{MSISDN}{Mobile Station International Subscriber Directory Number}
\newacronym{MSP}{MSP}{Mobile Service Provider}
\newacronym{MT}{MT}{Maximum Throughput}
\newacronym{NAS}{NAS}{Non Access Stratum}
\newacronym{NE}{NE}{Nash Equilibrium}
\newacronym{NR}{NR}{New Radio}
\newacronym{NTN}{NTN}{Non-Terrestrial Network}
\newacronym{NFV}{NFV}{Network Functions Virtualization}
\newacronym{NoSQL}{NoSQL}{Not Only SQL}
\newacronym{OAM}{OAM}{Operation, Administration and Management}
\newacronym{OFDM}{OFDM}{orthogonal frequency division multiplexing}
\newacronym{OFDMA}{OFDMA}{orthogonal frequency division multiple access}
\newacronym{ONF}{ONF}{open networking foundation}
\newacronym{ONOS}{ONOS}{Open Network Operating System}
\newacronym{OPEX}{OpEx}{operating expenditure}
\newacronym{OS}{OS}{operating system}
\newacronym{OTT}{OTT}{over-the-top}
\newacronym{PCI}{PCI}{Physical Cell Identity}
\newacronym{PCRF}{PCRF}{Policy and Charging Rules Function}
\newacronym{PDF}{PDF}{Probability Distribution Function}
\newacronym{PDN}{PDN}{packet data network}
\newacronym{PDCP}{PDCP}{Packet Data Convergence Control}
\newacronym{PDU}{PDU}{Protocol Data Unit}
\newacronym{PF}{PF}{Proportional Fair}
\newacronym{PGW}{P-GW}{Packet Data Gateway}
\newacronym{PHY}{PHY}{physical layer}
\newacronym{PoC}{PoC}{Proof-of-Concept}
\newacronym{PPP}{PPP}{{P}oisson point process}
\newacronym{PTP}{PTP}{Precision Time Protocol}
\newacronym{QoE}{QoE}{quality-of-experience}
\newacronym{QoS}{QoS}{quality-of-service}
\newacronym{PSC}{PSC}{Primary Scrambling Code}
\newacronym{PSD}{PSD}{power spectral density}
\newacronym{RACH}{RACH}{random access channel}
\newacronym{RAN}{RAN}{Radio Access Network}
\newacronym{RAT}{RAT}{Radio Access Technology}
\newacronym{RB}{RB}{resource block}
\newacronym{RE}{RE}{range extension}
\newacronym{RF}{RF}{radio frequency}
\newacronym{RG}{RG}{rate guarantee}
\newacronym{RLC}{RLC}{Radio Link Controller}
\newacronym{RNC}{RNC}{Radio Network Controller}
\newacronym{RR}{RR}{Round Robin}
\newacronym{RRC}{RRC}{Radio Resource Control}
\newacronym{RRH}{RRH}{remote radio head}
\newacronym{RRU}{RRU}{Remote Radio Unit}
\newacronym{RRM}{RRM}{radio resource management}
\newacronym{RSI}{RSI}{RACH Root Sequence Index}
\newacronym{RSS}{RSS}{received signal strength}
\newacronym{RSSI}{RSSI}{received signal strength indicator}
\newacronym{RSRP}{RSRP}{reference signal received power}
\newacronym{RTT}{RTT}{Round Trip Time}
\newacronym{SAC}{SAC}{service area code}
\newacronym{SAW}{SAW}{simple additive weighting}
\newacronym{SC-FDMA}{SC-FDMA}{single carrier frequency division multiple access}
\newacronym{SCN}{SCN}{small cell network}
\newacronym{SCTP}{SCTP}{Stream Control Transmission Protocol}
\newacronym{SDN}{SDN}{Software-Defined Networking}
\newacronym{SDMN}{SDMN}{Software Defined Mobile Network}
\newacronym{SDU}{SDU}{Service Data Unit}
\newacronym{SecGW}{SecGW}{Security Gateway}
\newacronym{SGSN}{SGCN}{Serving GPRS Support Node}
\newacronym{SGW}{S-GW}{Serving Gateway}
\newacronym{SHARING}{SHARING}{Self-organized Heterogeneous Advanced RadIo Networks Generation}
\newacronym{SINR}{SINR}{signal-to-interference-plus-noise ratio}
\newacronym{SISO}{SISO}{single-input single-output}
\newacronym{SSID}{SSID}{Service Set Identification}
\newacronym{ST}{ST}{Standart Multi-User TOPSIS}
\newacronym{STBCs}{STBCs}{space-time block codes}
\newacronym{SyncE}{SyncE}{Synchronous Ethernet}
\newacronym{TB}{TB}{Transport Block}
\newacronym{TCP}{TCP}{Transport Control Protocol}
\newacronym{TDMA}{TDMA}{Time Division Multiple Access}
\newacronym{TEID}{TEID}{tunnel endpoint identifier}
\newacronym{TOPSIS}{TOPSIS}{Total Order Preference By Similarity to the Ideal Solution}
\newacronym{TTI}{TTI}{transmission time interval}
\newacronym{UDP}{UDP}{User Datagram Protocol}
\newacronym{UE}{UE}{user equipment}
\newacronym{UL}{UL}{Uplink}
\newacronym{UP}{UP}{User Plane}
\newacronym{UMTS}{UMTS}{Universal Mobile Telecommunications Service}
\newacronym{URLLC}{URLLC}{Ultra-reliable low latency communications}
\newacronym{VoIP}{VoIP}{voice over IP}
\newacronym{VPN}{VPN}{virtual private network}
\newacronym{VSAT}{VSAT}{Very Small Aperture Terminal}
\newacronym{W-CDMA}{W-CDMA}{Wideband Code Division Multiple Access}
\newacronym{WiFi}{WiFi}{Wireless Fidelity}
\newacronym{Wi-Fi}{Wi-Fi}{Wireless Fidelity}
\newacronym{WiMAX}{WiMAX}{Worldwide Interoperability for Microwave Access}
\newacronym{WLAN}{WLAN}{Wireless Local Area Network}
\newacronym{WMC}{WMC}{weighted Markov chain}
\newacronym{ZF}{ZF}{zero-forcing}
\newacronym{MNO}{MNO}{Mobile Network Operator}
\newacronym{SON}{SON}{Self Organizing Network}
\newacronym{ANR}{ANR}{Automatic Neighbor Relation}
\newacronym{MRO}{MRO}{Mobility Robustness Optimizer}
\newacronym{MLB}{MLB}{Mobility Load Balancing}
\newacronym{CQO}{CQO}{cell quality offset}

\begin{document}


\title{On the Impact of Satellite Communications over Mobile Networks: An Experimental Analysis}

\author{Engin~Zeydan,~\IEEEmembership{Member,~IEEE,}
        Yekta~Turk,~\IEEEmembership{Member,~IEEE,}
\thanks{E. Zeydan is with  Centre Technologic de Telecomunicacions de Catalunya, Castelldefels, Barcelona, Spain, 08860. Email: engin.zeydan@cttc.cat}
\thanks{Y. Turk is a Mobile Network Architect based in Istanbul, Turkey. 34746. E-mail: yektaturk@ieee.org}%

}



\maketitle

\begin{abstract}
Future telecommunication systems are expected to co-exist with different backhauling nodes such as terrestrial or satellite systems. Satellite connectivity can add flexibility to backhauling networks and provide an alternative route for transmission. This paper presents experimental comparisons of satellite and terrestrial based backhaul networks and evaluates their performances in terms of different Key Performance Indicators (KPIs) including Channel Quality Index (CQI), Modulation Coding Scheme (MCS) index, Downlink (DL) throughput, Frame Usage (FU) ratio and number of Resource Block (RB) utilization. Our experimental satellite network system uses a real satellite-based backhaul deployment and works in Ka band. As a benchmark, we  compare our system with terrestrial network with regular cellular backhaul connection.  Our experiments reveal three main observations: The first observation is that problems with FU ratio and number of RB utilization exist in satellite eNodeB even though a single test user equipment (UE) with high CQI and MCS index values is connected. The second observation is that in satellite link relatively low numbers of Protocol Data Units (PDUs) are generated at Radio Link Controller (RLC) layer compared to the Packet Data Convergence Control (PDCP) layer. Finally, our third observation concludes that  the excessive existence of PDCP PDUs can be due to utilization of General Packet Radio Service (GPRS) Tunneling Protocol-User Plane (GTP-U) accelerator where an optimal balance between the caching size and the number of UEs using satellite eNodeB is needed. Hence, the existence of a trade-off between the supported number of UEs using satellite link and the GTP-U acceleration rate is also revealed with our experimental results.
\end{abstract}



\IEEEpeerreviewmaketitle

\section{Introduction}

\ac{LTE-A} is already putting a heavy burden on backhaul networks and various advanced techniques are proposed  to improve microwave backhaul links including \ac{AMC}, interference mitigation/cancellation techniques, higher order modulations, packet header compression, frequency diversity and \ac{MIMO}. This bottleneck problem of backhaul links gains more momentum as number of small cells inside \glspl{MNO}' network infrastructure is starting to soar. To carry this enormous amount of data traffic  generated inside cells by end-users and \ac{IoT} devices, in addition to \ac{RAN} level enhancements backhaul links have  to be redesigned as well. Otherwise, backhaul links will soon be the bottleneck that will put whole proper operations of end-to-end system into trouble. For these reasons, designing backhaul links is the next critical issue for \ac{5G} networks.  For backhaul connection, various communication mediums can be considered as candidates such as microwave radio, copper \ac{DSL}, optical fiber, \ac{mmWave} or satellite. The choice may depend on many factors including cost, performance, bandwidth demand, capacity and often many more. In particular, satellite point-to-point links that can exploit the benefits of satellite networks can be considered for reliable backhauling without interference to other cells or access links. Hence, satellites can play a key role for reliable service delivery in \ac{5G} networks and are already included in several \ac{5G} use cases~\cite{Jia2016}.

On the other hand, providing ubiquitous coverage is one of the major requirements that needs to be satisfied within \ac{5G} networks. Next generation \ac{5G} wireless solutions are expected to embrace both satellite and cellular solutions~\cite{jou2018architecture, boero2018satellite}. Due to cost per bit reductions of current state-of-the-art satellite technologies, \glspl{MNO} have started to select satellite technology as a promising backhaul solution inside their cellular infrastructure. In fact, satellite networks can provide \glspl{MNO} the opportunity to extend their coverage range in rural or remote areas of the country where existing infrastructure is not available (due to challenging topology of the geography) or limited in terms of network capacity. Hence, providing backhaul using satellite links in cellular networks can be a practical solution to provide low cost and timely delivery of connectivity services for hard to serve areas (such as mountains, islands, etc). For example, seamless integration and convergence with \ac{5G} terrestrial systems can support various vertical use cases and drive growth in sectors such as backhaul and trunking, transportation (aero, land, maritime) with mobile communications, media and entertainment with broadband services and public safety during disaster relief and emergency response situations.

Satellites can also complement next generation \ac{5G} terrestrial systems and provide substantial economic and societal benefits. For example, \glspl{MNO} can have the option of not building a new cellular backhaul infrastructure in rural or remote areas depending on return-of-investment over infrastructure. The usage of satellite network for backhaul can reduce the infrastructure investment cost while providing coverage to large-geographical areas of the country.

In light of these, it is inevitable that satellite networks will integrate with other networks including \ac{5G} cellular networks.  Support for satellite communications is considered to be an essential capability of the \ac{5G} technology. Moreover, satellites can be used to support key usage scenarios of \ac{5G} including \ac{eMBB}, \ac{mMTC} and \ac{URLLC}. For example, satellites can be used to carry high bandwidth \ac{HD} content via \glspl{HTS} in \ac{GEO}, \ac{MEO} and \ac{LEO} in \ac{eMBB} scenarios, can scale to support future \ac{IoT} communications in \ac{mMTC} scenarios and can play a role in low latency by delivering the same content to mobile \glspl{BS} or multicasting the content to caches of individual cells in \ac{URLLC} scenarios. For \ac{URLLC} applications, GSMA Intelligence report in~\cite{GSMA} states that the content for services that require less than 1 msec delay time should have all content served from a physical location that are very close to \ac{UE}, possibly at the base of every cell.  In fact, applications such as video on demand streaming, virtual/augmented reality or tactile Internet necessitate the requirement to move network capacities and capabilities to the edge~\cite{zeydan2016big}. Hence when the transition to \ac{5G} occurs, these new contents that require low latency need to be moved to the edge which will also require many new locations. In these situations, satellites can help \ac{5G} networks to meet their sub-1ms latency requirements by delivering commonly accessed content to mobile \glspl{BS}.  This is even possible even if no fiber connection at the dedicated site is available. According to Electronics Communication Committee  (ECC) report 280 approved on April 2018~\cite{ECC}, satellite can play an important role to connect and update large number of edge servers that next generation mobile networks will require.  Therefore, by multi-casting content to caches that are located at individual cells,  satellite multicast can become a viable option when \glspl{CDN} become ultimately densified.  However, the success of satellites also depends on their capabilities to provide cost/bit reduction compared to terrestrial systems as well as adequate throughput improvements to provide \ac{5G} services for backup and offloading purposes. Moreover, large delays experienced by satellite links are one of the significant impediments to use this technology in traditional cellular networks, either in \ac{LTE} currently or \ac{5G} in future.

\section{Motivation and Related Work}

Satellite communication can be carried out by four different satellite categories namely, \ac{HEO}, \ac{LEO}, \ac{MEO} and \ac{GEO} satellites that are orbiting at different altitudes around the Earth. In general, \ac{GEO} and \ac{MEO} satellites are used for communication purposes, especially for satellite backhaul applications.

Several works exist in the literature that study satellite integration with cellular networks~\cite{giambene2018satellite, breiling2014lte,guidotti2017lte,ai2018ka,kapovits2018satellite,guidotti2018architectures,bastia2009lte,zangar2018leveraging}. Different satellite-\ac{5G} integration use cases as well as the latest initiatives and challenges in future \ac{5G} terrestrial and satellite integration are summarized in~\cite{giambene2018satellite}. In~\cite{giambene2018satellite}, the authors have also investigated the impact of impairments in a typical satellite channel in \ac{LTE} waveform design as well as in L1 and L2 procedures.  The paper in~\cite{breiling2014lte} uses a real-time simulator for satellite backhauling of moving \glspl{eNodeB} where \ac{LTE} network and satellite links are emulated.  The authors in~\cite{araniti2016multimedia} are proposing a new radio resource management algorithm for multimedia content distribution over satellite networks.

The authors in~\cite{kapovits2018satellite} study the integration of satellite and terrestrial communication networks and validate such integration with a testbed including a satellite emulator for backhaul support. The paper in~\cite{guidotti2018architectures} studies key technical challenges (mostly related to PHY/MAC layers) as well as architectures for incorporation of satellites into \ac{5G} systems.  The paper in~\cite{bastia2009lte} presents several enabling techniques to  reuse the existing terrestrial air interface for transmission over satellite links. The paper in~\cite{zangar2018leveraging} proposes a \ac{DL} scheduling strategy in an integrated terrestrial‐-satellite network to enhance spectrum efficiency, fairness and capacity. The authors in~\cite{shaat2018integrated} have reviewed the benefits of integration of satellite and terrestrial links to provide a network with wireless backhaul. The paper in~\cite{Niephaus} identifies the technical challenges associated with the convergence of satellite and terrestrial networks to provide \ac{QoS} similar to  high bandwidth terrestrial
networks for end-users. Performance evaluation and research challenges of multi-satellite relay systems with cooperative transmission in \ac{TDMA}-based architecture are given in~\cite{bai2018multi}. For delivering TV services, Rajeev Kumar et al. in~\cite{kumar2018wilitv} investigate the usage of different wireless links (including satellite, WiFi, and \ac{LTE}/\ac{5G} \ac{mmWave}) to improve TV distribution penetrations.

\begin{figure*}[t]
\centering
\includegraphics[width=0.9\linewidth]{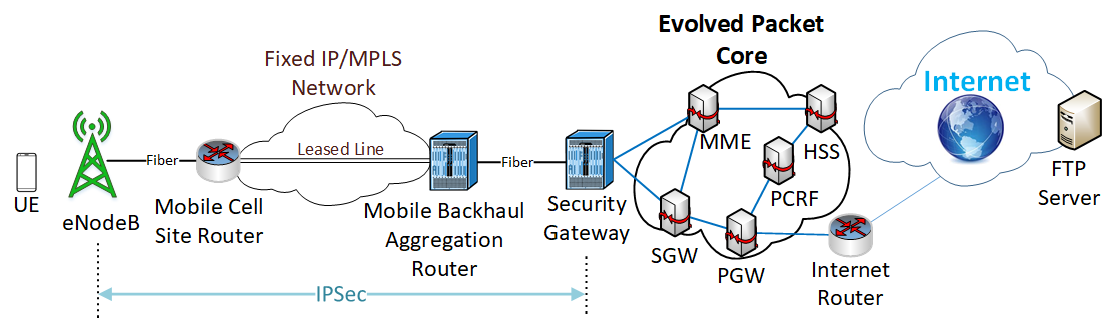}
\caption{General architecture of the terrestrial network.}
\label{satellite_simplified}
\end{figure*}

Standardization bodies including \ac{3GPP}~\cite{3GPPTR38811}, \ac{ETSI}~\cite{ETSI}, \ac{ITU}~\cite{ITUSatellite} as well as joint initiatives such as \ac{5G PPP}~\cite{5GPPVisison} have also considered satellite networks in conjunction with terrestrial communication systems. For instance, \ac{3GPP} Release $15$ has recently drafted a technical study on \ac{NR} to support \glspl{NTN}~\cite{3GPPTR38811}. \ac{EC} funded \ac{5G PPP} projects have also been launched within H2020 framework that focus on satellite communication. SANSA (Shared Access terrestrial-satellite backhaul Network enabled by Smart Antennas) project\footnote{https://sansa-h2020.eu/} investigates the utilization of extended Ka band for backhaul operations to improve spectrum efficiency. SaT5G (Satellite and Terrestrial Network for 5G)\footnote{http://sat5g-project.eu/} project focuses on  plug-and-play integration of satellite communications into \ac{5G} network for \ac{eMBB} use case. ESA ARTES SATis5 (Demonstrator for Satellite-Terrestrial Integration in \ac{5G} Context)\footnote{https://artes.esa.int/projects/satis5} project builds a large \ac{PoC} testbed to enable satellite-terrestrial convergence into \ac{5G} context focusing on \ac{eMBB} and \ac{mMTC} scenarios.

Additionally, various works investigate satellite communication in Ka band~\cite{guidotti2017lte, ai2018ka, zhao2018beam, hasan2016ka}. The authors in~\cite{guidotti2017lte} focus on extending the \ac{LTE} broadband service using a mega-constellation of \ac{LEO} satellites that are deployed in Ka-band. An \ac{UL} \ac{SINR} probabilistic model for Ka band in \glspl{HTS} is studied in~\cite{ai2018ka}.   The authors in~\cite{zhao2018beam} focus on beam tracking methodology in Ka-band for UAV-satellite communication systems. The paper in~\cite{hasan2016ka} provides an overview of advancements in many satellite based communication systems utilizing Ka band frequency band.

In operational networks, to compensate against the excessive latency existent in satellite links, \ac{GTP}-User Plane (GTP-U) accelerator is considered to be a viable solution by satellite modem vendors~\cite{GiLAT,Modulo}. The patent in~\cite{ahluwalia2016acceleration} has applied acceleration function to \ac{GTP} \glspl{PDU} for traffic flows over satellite links via utilization of caching, pre-fetching and web acceleration methods. \ac{GTP}-U accelerator is mainly useful for satellite links to mitigate the effect of long propagation delays that signal experiences.

Although most of the above papers have investigated cellular-satellite integration aspects, some limitations of the considered scenarios exist. For example, some related works either investigate performance over satellite emulators testbed implementations or work on virtualized environments of real-time simulators~\cite{breiling2014lte,kapovits2018satellite}. These results lack relevant realistic requirements including the combined effect of \ac{LTE} radio link with satellite-based backhaul connection to \ac{EPC}.  The aim of this work is to show how performance of satellite-based backhaul solution behaves in a real-time operational network and in comparison with regular terrestrial network. An important contribution of this paper is to complement literature works on integrated satellite-terrestrial networks  by presenting an end-to-end complete and realistic validation results of the considered satellite-based backhaul system. Regarding that aspect, our proposed analysis works around the deployment of a satellite-based backhaul architecture, analysis over a wide range of \ac{KPI} measurements including number of \glspl{RB} utilization, \glspl{FU} ratios, \ac{CQI}, \ac{MCS} index, \ac{PDCP} throughput, number of \glspl{PDU} in \ac{RLC} and \ac{PDCP} layers and \ac{MIMO} \ac{TB} usage and comparisons with an operational terrestrial cellular network. More specifically, a eNodeB that connects a test \ac{UE} to \ac{EPC} via satellite-based backhaul is considered and the impact of satellite link latency on resource allocation  schemes of satellite eNodeB is investigated.


Our experimental evaluations  are conducted at minutes-level intervals which provide a fine grained characterization of the considered satellite-based backhaul deployment scenario.  The analysis results reveal observed problems and trade-offs when \glspl{KPI} of experimental satellite-based backhaul and terrestrial networks are compared. All these considerations have allowed us to understand the behaviour satellite-based backhaul architecture under real operating conditions. Our contributions in this paper can be summarized as follows:

\begin{itemize}
    \item proposing a satellite-based backhaul architecture for cellular networks and performing a real-world experimentation over satellite link that utilizes Ka band,
    \item investigating satellite-based backhaul architecture's performance impacts on radio level \glspl{KPI} and their comparisons with terrestrial cellular network's \glspl{KPI},
    \item revealing three distinct observations based on existence of \ac{FU} ratio and \ac{RB} utilization problems, excessive  number of \ac{PDCP} \glspl{PDU} and the trade-off between the support for higher number of \glspl{UE} and \ac{GTP}-U acceleration rate using the satellite-based backhaul link.
\end{itemize}

The structure of the paper is as follows: Section~\ref{systemmodel} presents system model, concepts and the proposed architecture.  Section~\ref{experiment} presents the experimental analysis results using \ac{KPI} measurements from both satellite and terrestrial eNodeBs for comparison purposes.  The paper ends with conclusions in Section~\ref{conclusions}.

\section{System Model and Architecture}
\label{systemmodel}

Fig.~\ref{satellite_simplified} shows end-to-end system architecture of the utilized testbed for the terrestrial network site. This traditional mobile network consists of \ac{RAN}, transport and core networks.  In \ac{LTE}, the access network is called \ac{E-UTRAN} and the core network is \ac{EPC} with all \ac{IP}-based connection. \ac{E-UTRAN} is composed of eNodeBs, \glspl{UE} whereas \ac{EPC} includes \ac{PGW}, \ac{SGW}, \ac{PCRF}, \ac{HSS} and \ac{MME}.

\begin{figure*}[t]
\centering
\includegraphics[width=0.8\linewidth]{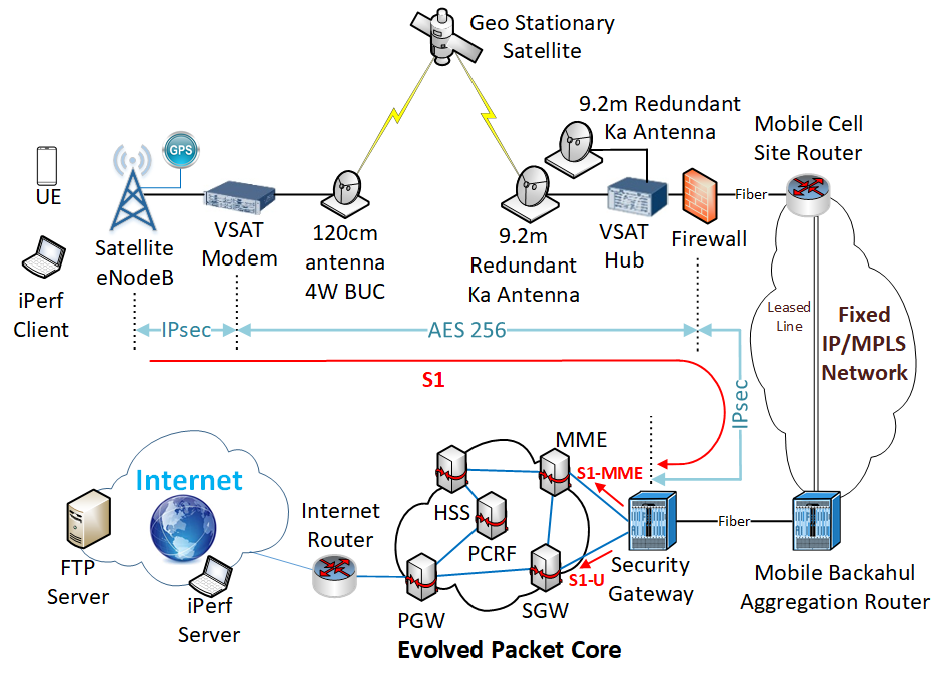}
\caption{General architecture of the satellite-based backhaul network.}
\label{satellite_arch}
\end{figure*}

\subsection{General architecture of the satellite-based backhaul network}

Fig.~\ref{satellite_arch} represents a high-level architecture and key components of a \ac{GEO} satellite-based backhaul for mobile network communication. We consider a \ac{GEO} satellite that can provide satellite service to mobile network.  In this setup, S1 interface between eNodeb and \ac{EPC} is transported via satellite. \ac{LTE} supports satellite transmission over S1 interface. \ac{VSAT} modem is connected to the eNodeB with an Ethernet cable. \ac{VSAT} hub and modem communicate over \ac{GEO} satellite. \ac{VSAT} hub and \ac{VSAT} modem perform optimization to the S1 traffic as well as modulation of signal to be transmitted over the satellite link.  In addition, \ac{VSAT} modem is capable of both \ac{IPsec} setup and \ac{GTP}-U acceleration. A $120$cm antenna is deployed on the site for satellite access. Two redundant $9.2$m Ka antennas, one \ac{VSAT} hub and one firewall are located in ground station. \ac{VSAT} hub is connected to  firewall via Ethernet.  Between \ac{VSAT} modem and firewall, \ac{AES}-256 encryption is used to encrypt traffic since \ac{IPsec} usage is not feasible over satellite networks. The firewall is used for encryption of the satellite traffic and establishment of a new \ac{IPsec} tunnel with the \ac{SecGW} in the mobile core network. In \glspl{MNO}' networks, \glspl{CSR} are installed at eNodeB sites to transmit traffic to the \ac{IP}/\ac{MPLS} backhaul network. However due to this topology where eNodeB S1 traffic is transmitted over satellite, mobile \ac{CSR} is installed on the ground station where \ac{VSAT} hub is located. The \ac{CSR} is connected to a mobile \ac{MBAR} via a leased line over a carrier of the fixed network operator. Links from many \glspl{CSR} are collected in the \ac{MBAR} and forwarded to the \ac{LTE} \ac{SecGW}. The purpose of the \ac{SecGW} is to terminate \ac{IPsec} tunnel set up and (d)encrypt S1 traffic. \ac{SecGW} is connected to \ac{EPC} and \ac{EPC} is connected to the Internet via a high-capacity Internet router as given in Fig.~\ref{satellite_arch}. \ac{GTP}-U acceleration method is applied mutually at both \ac{VSAT} modem and \ac{VSAT} hub~\cite{GiLAT}.

\subsection{Ka band utilization for satellite-based backhaul}

\glspl{HTS} are characterized by many small beams with high gains that also allow multiple frequency re-use. It can provide high capacity communication with reduced costs~\cite{fenech2015high}.  Ka band frequency utilization is getting common in \glspl{HTS}~\cite{hasan2016ka}. In fact, classical frequency bands such as C and Ku-band are getting congested due to increasing number of satellite communication systems. For this reason, utilization of Ka bands can provide higher bandwidths compared to L-/S- bands and is more suitable for broadband services due to high frequency usage. Moreover, higher frequencies avoid the interference with terrestrial communication systems, enable the reduction of system components both in ground and space segments and enable higher antenna gains and directivity. However, as frequency increases small perturbations on the atmosphere (e.g. rain attenuation) can have high impact on link quality of the propagating waves. This affects end-users' experienced \ac{QoS}. For this reason, it is critical to observe the effect of Ka band utilization over satellite links with experimental trials and compare various \glspl{KPI} with legacy terrestrial networks.

\section{Experimental Results}
\label{experiment}

In  this section, our focus is to obtain empirical assessments of the satellite-based backhaul set-up and compare it with a  terrestrial 4G cellular mobile radio communication network using different \glspl{KPI}. Observations of different \glspl{KPI} can yield insights into identifying trade-offs under actual network deployment conditions.  For this reason, a real-time prototype of satellite-based backhaul is deployed and tested over the infrastructure of the operator.

\begin{figure}[t]
\centering
\includegraphics[width=1.0\linewidth]{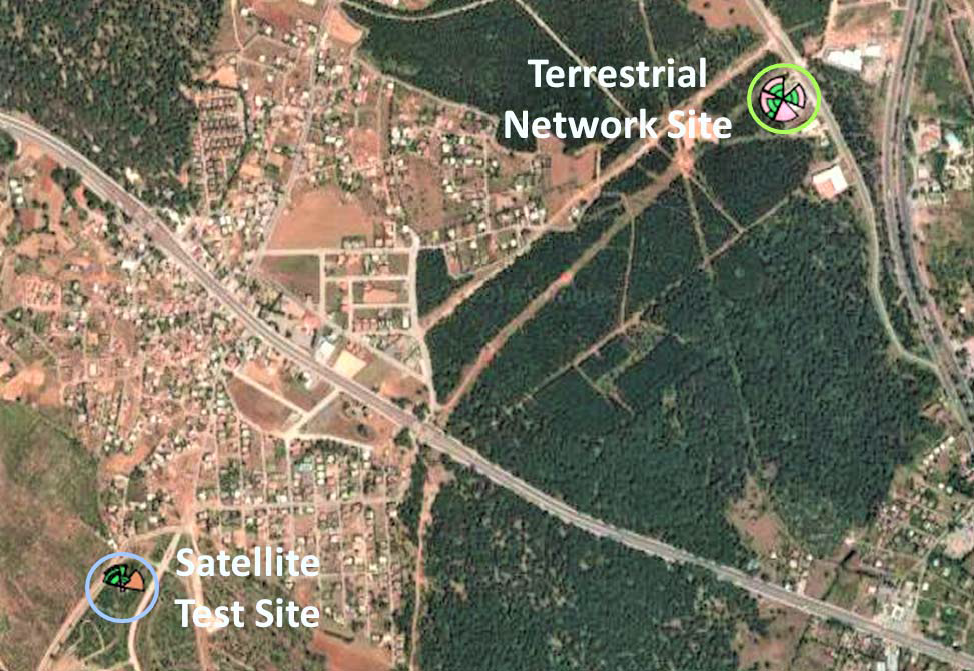}
\caption{Experimental test site locations of terrestrial and satellite networks.}
\label{satellite_map}
\end{figure}

In the presented work, the deployed satellite-based backhaul system constitutes a single satellite eNodeB where a single test \ac{UE} is connected to an \ac{EPC} for end-to-end system implementation. Fig.~\ref{satellite_map} shows experimental test site locations of the utilized terrestrial and satellite networks. The \ac{KPI} performance differences between a satellite and terrestrial-based backhaul mobile system enable us to evaluate their comparative performances and observe any potentially existing trade-offs.

\subsection{Satellite components and configurations}

Our satellite system is \ac{HTS} and  uses Ka band frequencies between $19.828.662$ kHz - $19.891.412$ kHz for \ac{DL} and $29.590.047$ kHz - $29.594.041$ kHz for \ac{UL}. Therefore, spectrum bandwidths are $70$ Mhz and $4$ Mhz for \ac{DL} and \ac{UL} respectively.

The \ac{DL} transmission is between the satellite and the eNodeB whereas \ac{UL} is vice-versa. Usually, dedicated bandwidth for \ac{DL} is higher than \ac{UL}. The total bandwidth is the summation of \ac{UL} and \ac{DL} bandwidths. In this paper, we  utilize a \ac{GEO} stationary satellite which is $35.786$ km away from Earth. Considering the GEO satellite altitude and its position in an equatorial plane, the distance between the GEO satellite and a point on the Earth's surface can be ranged between $35.786$ km (if the point is on the Equator) and about $40.000$ km (near to one of the pole). As a consequence,  one-way propagation delay of \ac{GEO} satellite transmission can be calculated to range between

\begin{equation}
    \frac{35,786 km}{3e5 km/s} = 119 ms \leq T =  \frac{d}{c} \leq  \frac{40,000 km}{3e5 km/s} = 133 ms.
\end{equation}

\noindent
Hence, the communication delay between two ground stations through the satellite is $238ms \leq 2T \leq 266ms$.

\begin{figure}[t]
\centering
\includegraphics[width=1.0\linewidth]{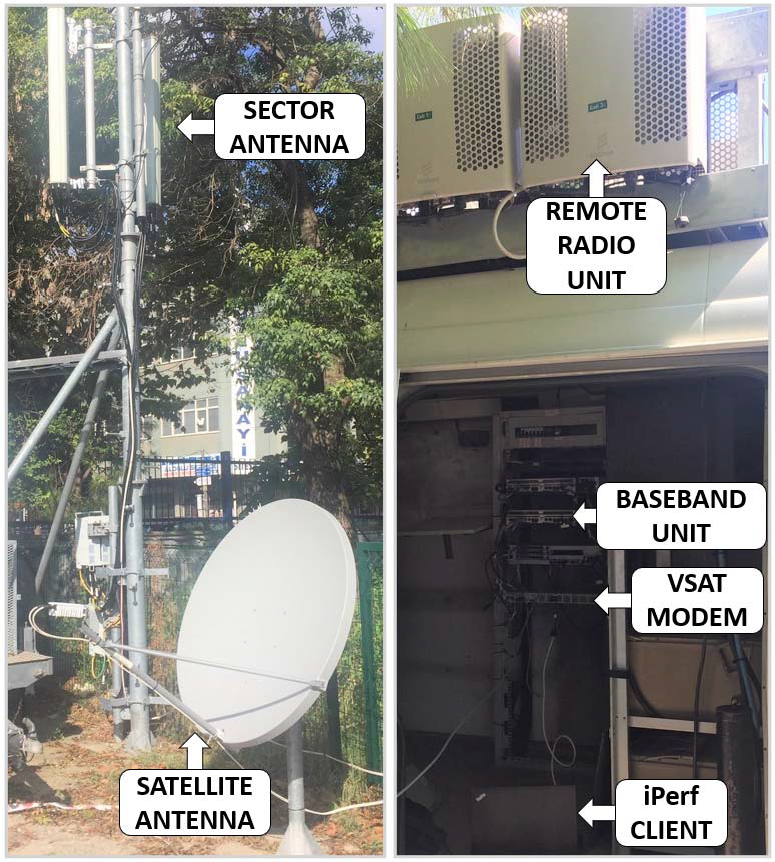}
\caption{Installed equipment and connections on the satellite test site. Left: Satellite and sector antennas.  Right: RRU, BBU, VSAT modem and Iperf client.}
\label{site_photo}
\end{figure}

\subsection{Testbed network functions}

For experimental end-to-end tests over the considered networks, we used a notebook with SIM card as an iPerf~\cite{iPerf} client and a test \ac{UE} with TEMS \cite{TEMS_datasheet} tool installed to collect radio-level \glspl{KPI}. A \ac{FTP} server that is located on the Internet is used to download a file via \ac{UE}. Same tests were conducted both in terrestrial and  satellite test sites. Iperf tool is used to generate a \ac{UDP} flow between the iPerf client and the iPerf server.  The principle objective of the experimental test system is to allow realistic evaluations of satellite-based backhaul deployment in a real-operator environment.  Fig.~\ref{site_photo} shows different types of network devices that are used during experimental trials. These components are:

\begin{enumerate}
    \item \textbf{Satellite antenna:} The antenna is fixed to the ground to combat against environmental effects after \ac{LOS} is adjusted towards the satellite.
    \item \textbf{Sector antenna:} A sector antenna is a directional antenna for outdoor environment that provides high gains. Sector antennas consist of an array of dipoles. The performance of these antennas depend on the size and shape of the reflector.
    \item \textbf{\ac{BBU}:} A \ac{BBU} is a unit that processes baseband and is connected to \ac{RRU} via optical fiber.
    \item \textbf{\ac{RRU}:} A \ac{RRU} performs \ac{RF} \ac{DL} and \ac{UL} channel processing. \ac{RRU} communicates with \ac{BBU} via a physical link and with wireless mobile devices via air interface.
    \item \textbf{\ac{VSAT} Modem:} An equipment that performs \ac{GTP}-U acceleration and selected security protocols (e.g. \ac{IPsec}, \ac{AES}-256). It is also compatible to operate with \ac{LTE}. It is connected to the satellite antenna and eNodeB.
    \item \textbf{iPerf Client:} iPerf works in a client/server mode. A client sends data to the server for testing purposes.
\end{enumerate}

Additionally, the deployment complies with the relevant features of \ac{LTE} standards.

\begin{table}[bt]
\centering
\begin{tabular}{|c|c|c|}
\hline
\textbf{\begin{tabular}[c]{@{}c@{}}Measured \\ Protocol\end{tabular}} & \textbf{\begin{tabular}[c]{@{}c@{}}Upload Test \\ Completion Time(s)\end{tabular}} & \textbf{\begin{tabular}[c]{@{}c@{}}Download Test \\ Completion Time(s)\end{tabular}} \\ \hline
TCP & 11.35 & 11.13 \\ \hline
UDP & 10.98 & 10.89 \\ \hline
\end{tabular}
\caption{Test completion results of iPerf measurements taken from the satellite eNodeB.}
\label{iperftest}
\end{table}

Table~\ref{iperftest} presents iPerf measurements' delay responses for \ac{UL} and \ac{DL} from/to the satellite eNodeB. Bandwidth setting for iPerf test is limited to $50$ Mbit/s to transfer approximately $60$ MBytes of data.  The results indicate that \ac{UDP} traffic has low completion time compared to \ac{TCP} traffic. In IPerf tests, it is seen that tests with \ac{TCP} were completed in both \ac{UL} and \ac{DL} by approximately $3$ percent later than \ac{UDP}. This demonstrates how \ac{TCP} is affecting the applications used by test \ac{UE} connected to satellite eNodeB. Differences between \ac{UL} and \ac{DL} test completion duration are caused by hardware differences between iPerf server and client. It is also related to the high number of simultaneous tests performed by the server.

In rest of the paper, \ac{PDF} of a \ac{KPI} represents the percentage of the specific \ac{KPI}'s measurements at a specific value whereas \ac{CDF} represents the percentage of that specific \ac{KPI}'s measurements that are at least as good as a specific value.

\subsection{Performance in Downlink}

\begin{figure*} [ht!]
\centering
\begin{subfigure}{0.8\textwidth}
  \centering
   \includegraphics[width=\linewidth]{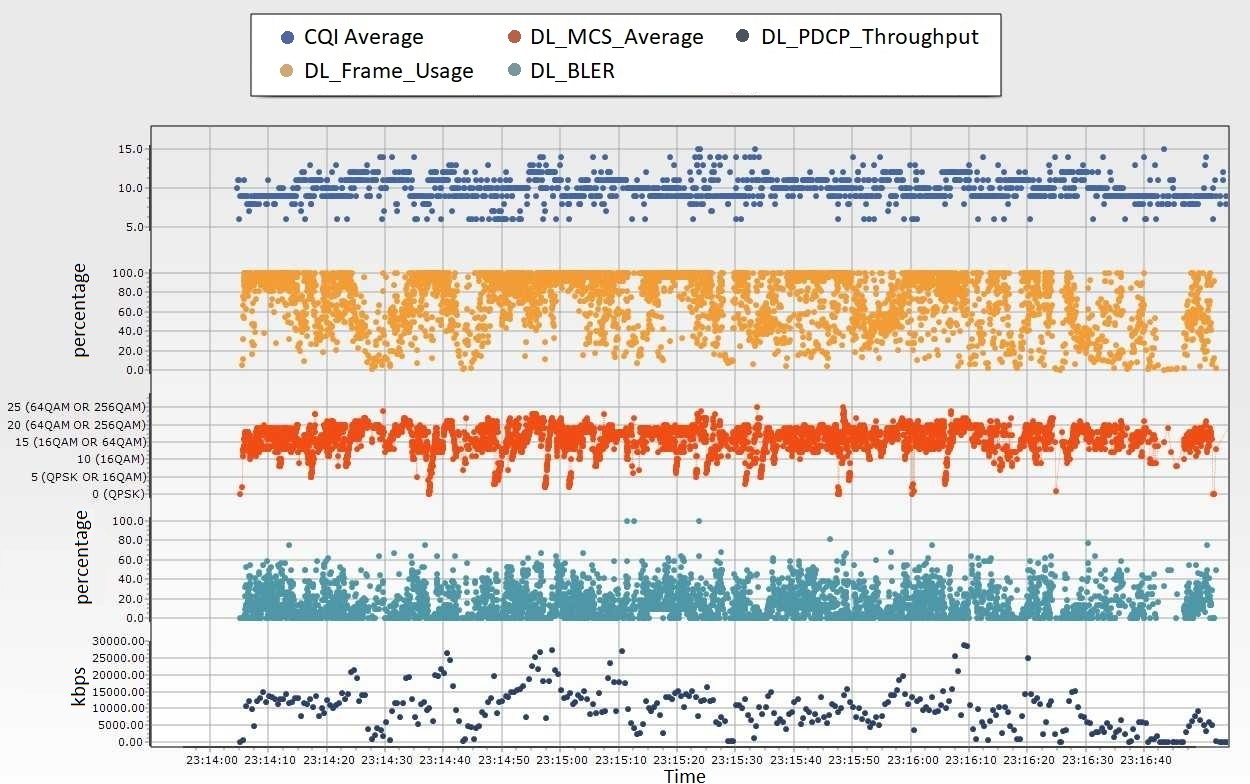}
  \caption{}
  \label{live1}
\end{subfigure} \\
\begin{subfigure}{0.8\textwidth}
  \centering
  \includegraphics[width=\linewidth]{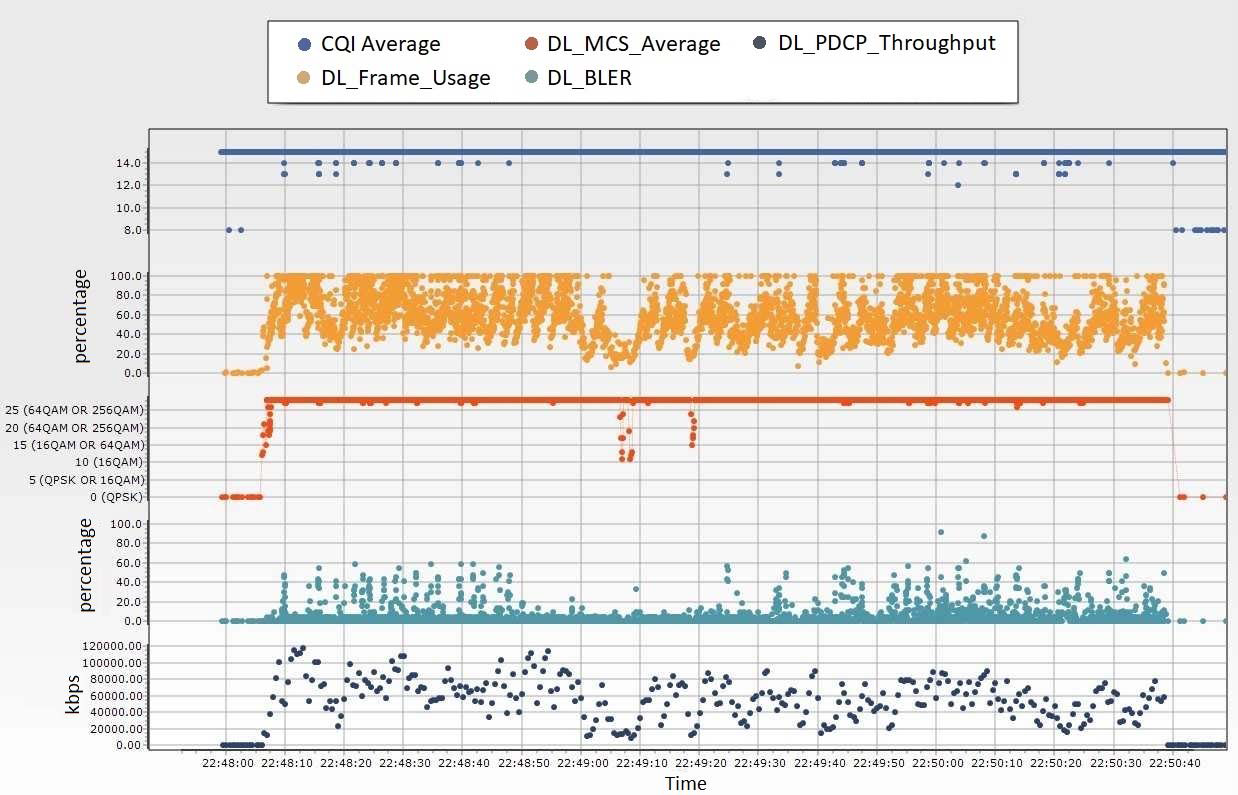}
  \caption{}
  \label{satellite_1}
\end{subfigure}\\
\caption{Radio Performance KPIs in DL (a) Terrestrial eNodeB (b) Satellite eNodeB [Figures are best viewed in color.]}
\label{tems_kpi_1}
\end{figure*}

\textbf{\ac{MCS} and \ac{CQI} values:} Top subfigures of Fig.~\ref{tems_kpi_1} marked with dark blue colors represent the average \ac{CQI} values for both terrestrial and satellite eNodeBs. We can observe that satellite eNodeBs' link quality is high and a mean \ac{CQI} value of $14.96$ is observed during the observation time interval. This indicates good \ac{RF} medium conditions in satellite-based backhaul network and there is no problem with \ac{RF} signal quality for satellite test \ac{UE}. On the other hand, terrestrial network's \ac{CQI} values fluctuate and radio conditions are slightly poor compared to satellite eNodeB with an observed mean \ac{CQI} value of $9.92$. Red colored marks in third row of Fig.~\ref{live1} and Fig.~\ref{satellite_1} show \ac{MCS} index values over the observation duration for terrestrial and satellite eNodeBs respectively. \ac{MCS} index values in terrestrial eNodeB are fluctuating between $0$ and $25$. This again validates that radio conditions are variable and worse than  satellite network. In comparison, satellite network \ac{MCS} index values are almost constant at maximum \ac{MCS} index value of $25$. This indicates usage of high modulation scheme in satellite eNodeB  with very small drop-offs at certain time intervals in \ac{DL}. Note that there are many real-\glspl{UE} connected to terrestrial eNodeB and only one test \ac{UE} in satellite eNodeB. This is also another effect that can have a negative impact on observed average \ac{CQI} and \ac{MCS} index values in terrestrial eNodeB.

\begin{figure} [t]
\centering
\begin{subfigure}{0.5\textwidth}
  \centering
   \includegraphics[width=\linewidth]{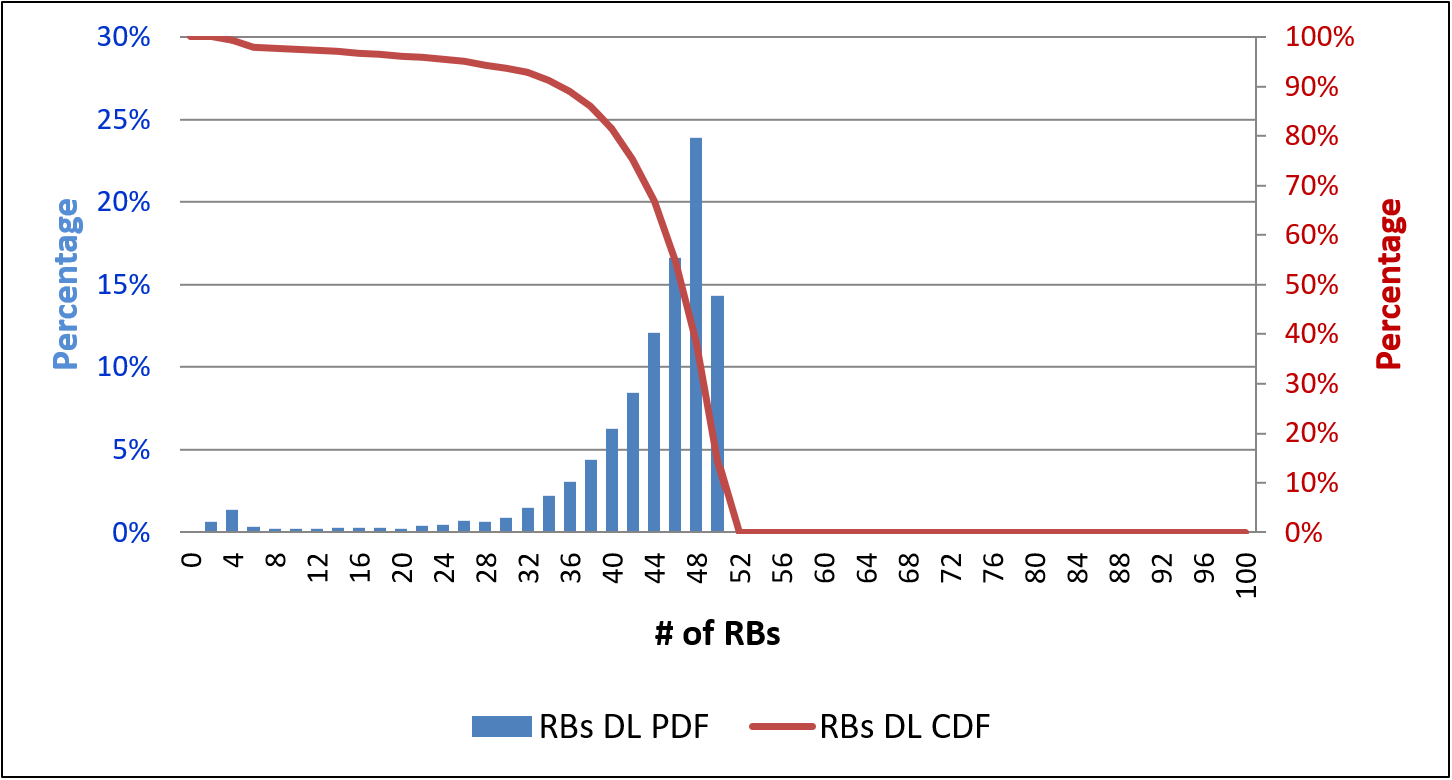}
  \caption{}
  \label{per3}
\end{subfigure} \\
\begin{subfigure}{0.5\textwidth}
  \centering
  \includegraphics[width=\linewidth]{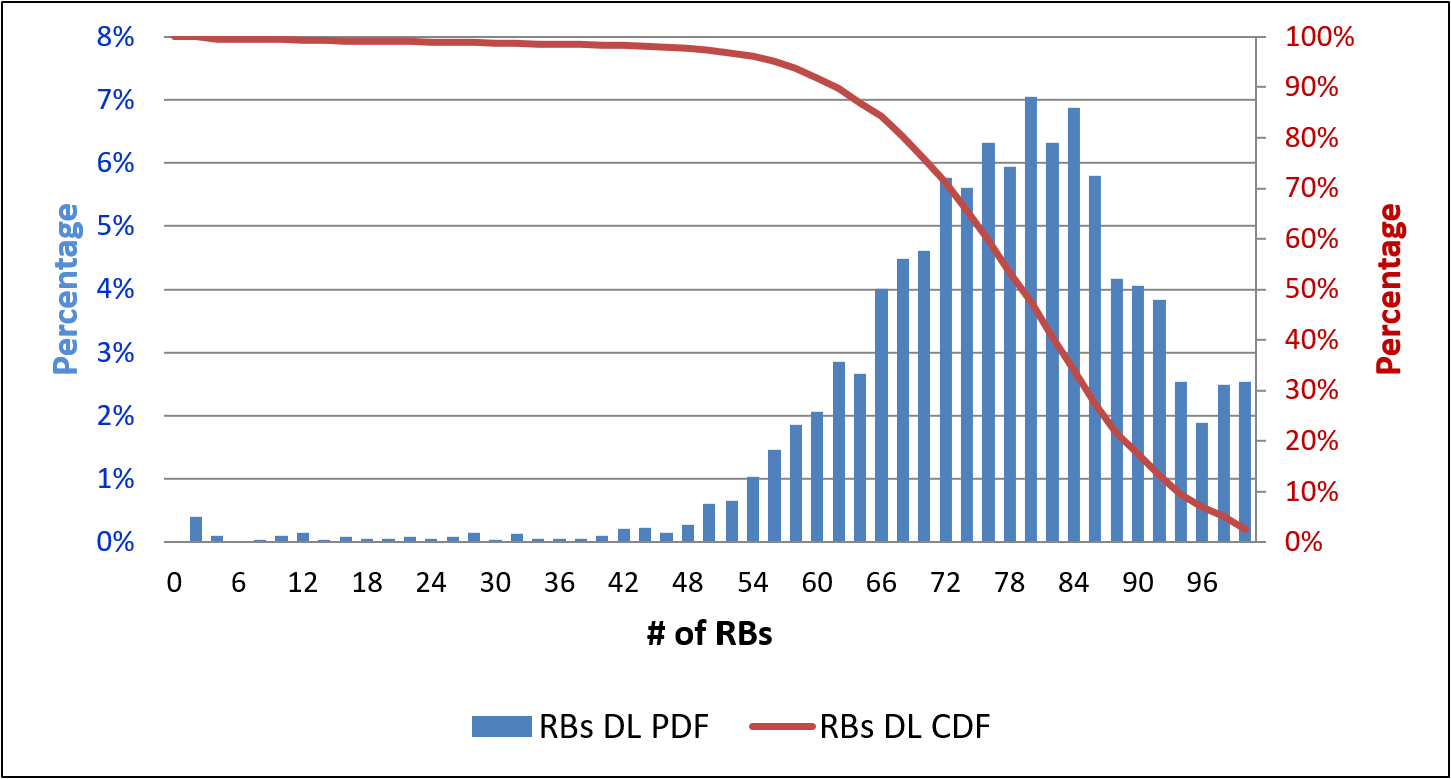}
  \caption{}
  \label{per4}
\end{subfigure}\\
\caption{PDF and CDF plots for DL number of RB utilization distributions in (a) Terrestrial eNodeB (b) Satellite eNodeB.}
\label{RB_Distribution}
\end{figure}

\begin{figure} [t]
\centering
   \includegraphics[width=\linewidth]{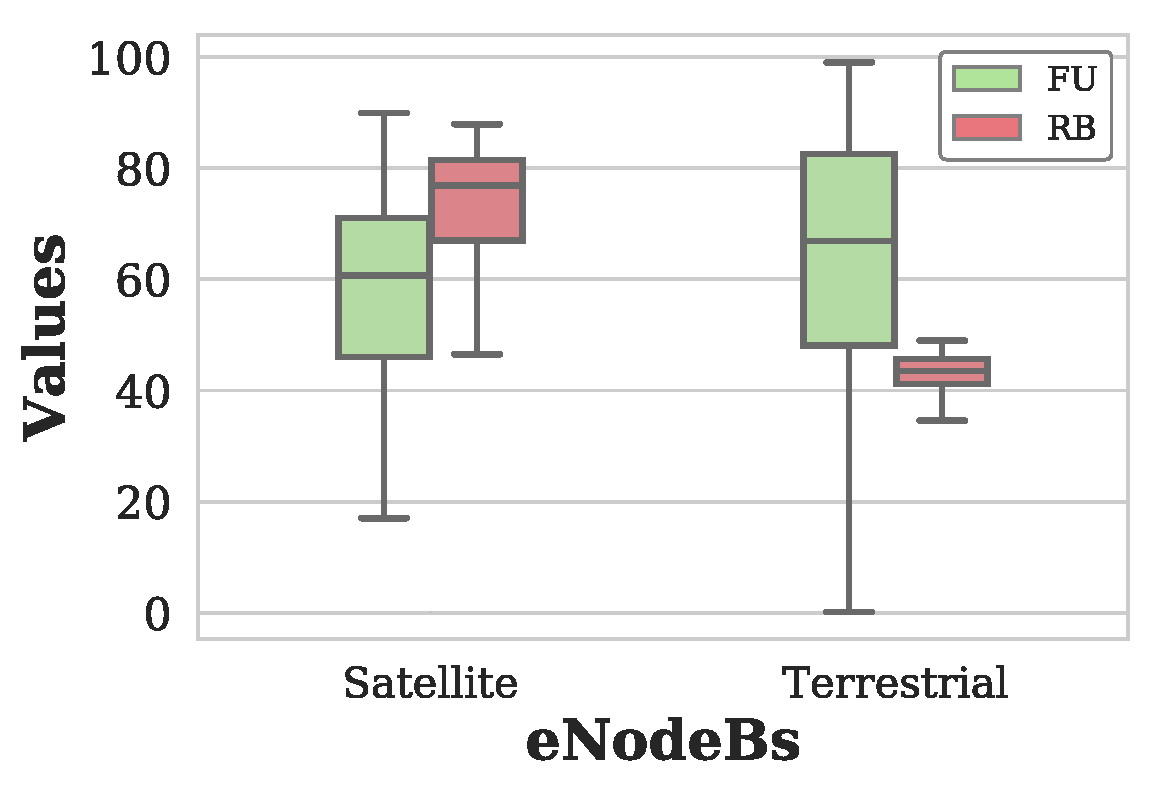}
\caption{Boxplot comparison values for number of RBs usage and the FU ratio (\%) in  DL for terrestrial and satellite eNodeBs.}
\label{fu_rb}
\end{figure}

\textbf{Number of \ac{RB} utilization distributions:}  Fig.~\ref{RB_Distribution} shows number of \ac{RB} utilization  in \ac{DL} for both terrestrial  and satellite eNodeBs.  From presented \ac{CDF} and \ac{PDF} plots, we can observe different distribution behaviours for number of \ac{RB} utilization in terrestrial and satellite eNodeBs. For example, in Fig.~\ref{RB_Distribution} (a) $82\%$ of number of \ac{RB} utilization are at least as good as  $40$ \glspl{RB} and in Fig.~\ref{RB_Distribution} (b)  $87\%$ of number of \ac{RB} utilization are at least as good as  $64$ \glspl{RB}. Similarly, Fig.~\ref{fu_rb} shows the boxplot for number of \glspl{RB} utilization in  \ac{DL} for both terrestrial and satellite eNodeBs. The median values for number of utilized \glspl{RB} for satellite and terrestrial eNodeB are $76.92$ and $43.53$ respectively. In terrestrial eNodeB, the variance of \ac{RB} is observed to be narrower than satellite eNodeB. Although \ac{RF} conditions seem relatively poor in terrestrial eNodeB, the scheduler utilizes \glspl{RB} as many as possible where up to $50$ \glspl{RB} are used for a given bandwidth of $10$ Mhz. In comparison, in satellite eNodeB, \ac{RB} utilization variance is larger.

\begin{figure} [t]
\centering
\begin{subfigure}{0.5\textwidth}
  \centering
   \includegraphics[width=\linewidth]{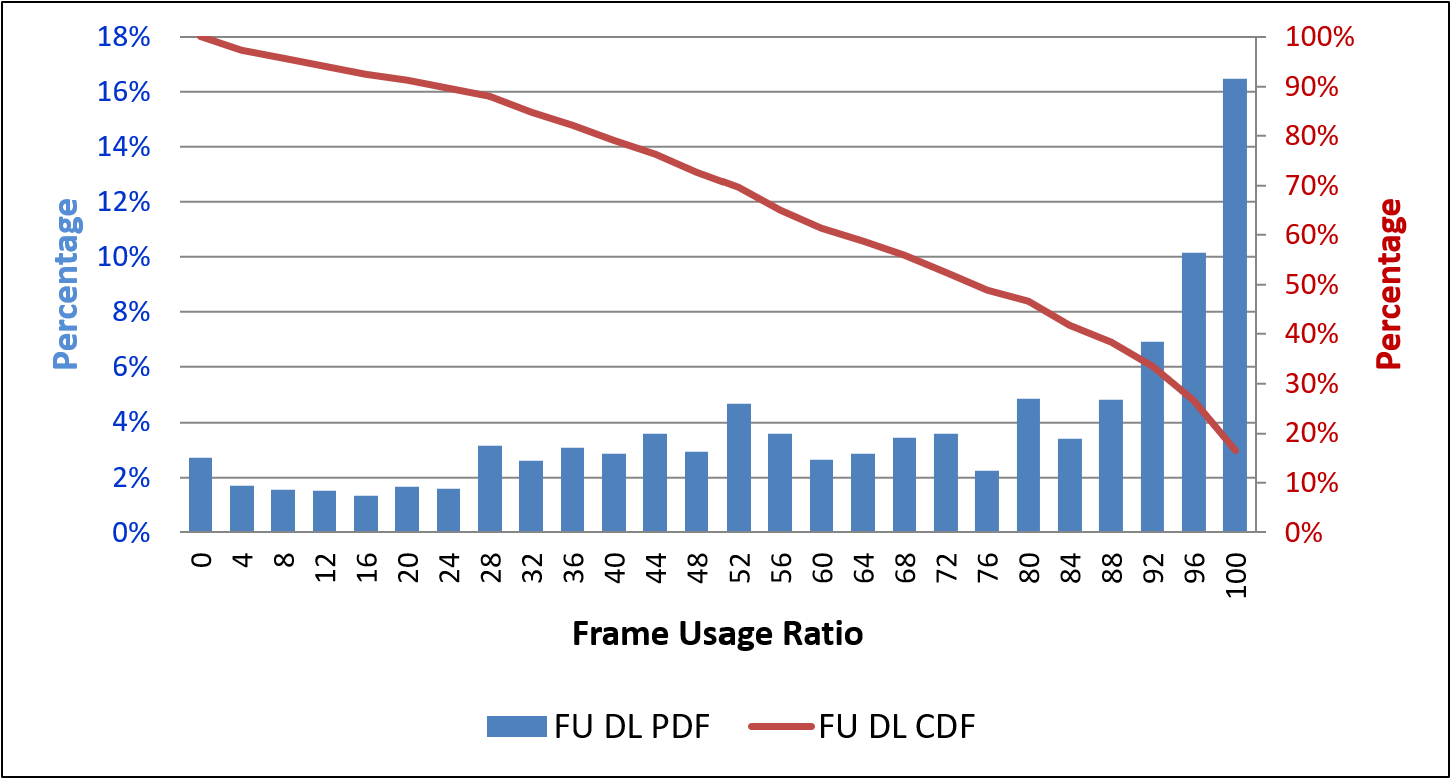}
  \caption{}
  \label{per33}
\end{subfigure} \\
\begin{subfigure}{0.5\textwidth}
  \centering
  \includegraphics[width=\linewidth]{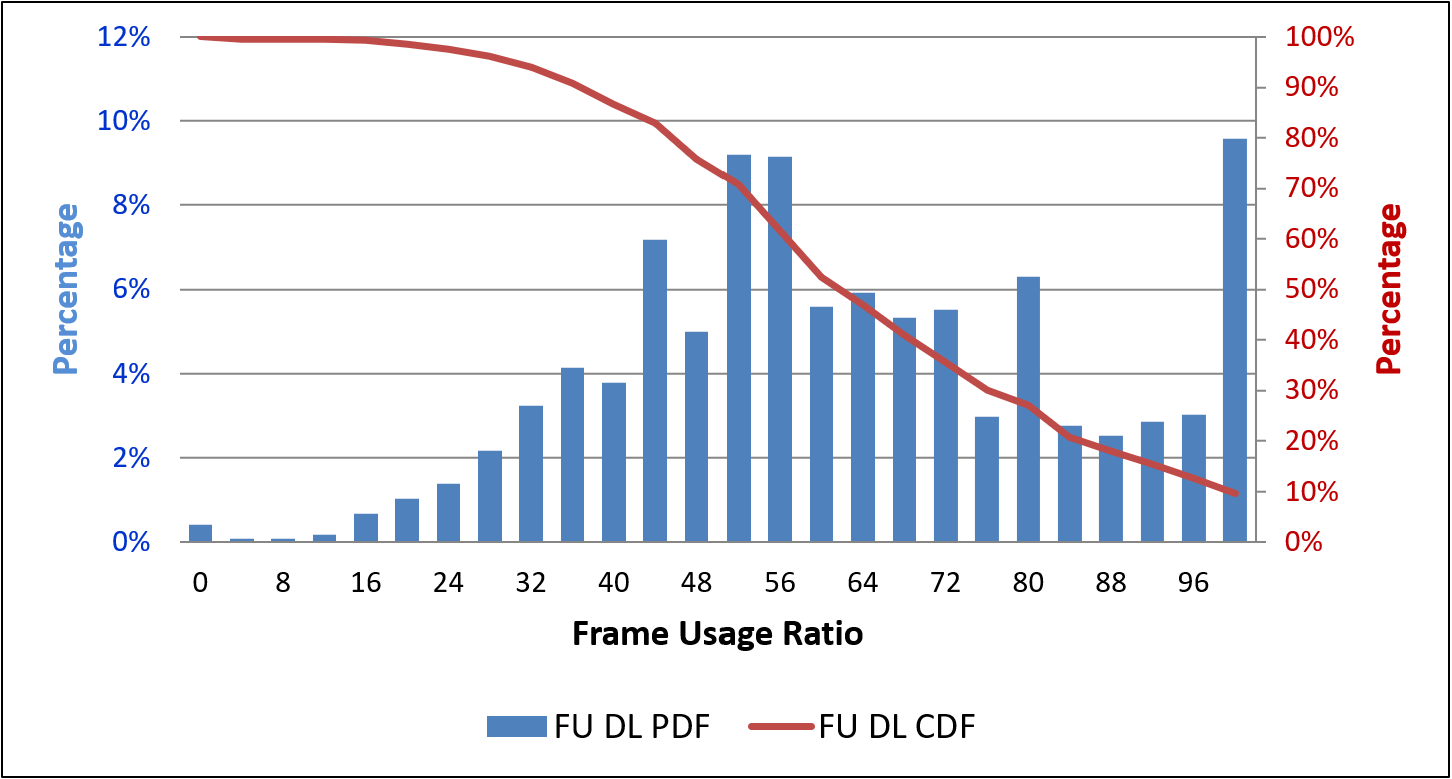}
  \caption{}
  \label{per44}
\end{subfigure}\\
\caption{PDF and CDF plots for DL FU ratio distributions in (a) Terrestrial eNodeB (b) Satellite eNodeB.}
\label{FU_Distribution}
\end{figure}

\textbf{\ac{FU} ratio distributions:}  Fig.~\ref{FU_Distribution} shows \ac{FU} ratios in \ac{DL} for both terrestrial and satellite eNodeBs.  \ac{CDF} and \ac{PDF} plots show the behaviour of \ac{FU} ratios distribution. For example, in Fig.~\ref{FU_Distribution} (a) $55\%$ of \ac{FU} ratio measurements are at least as good as  $68$ and in Fig.~\ref{FU_Distribution} (b)  $65\%$ of \ac{FU} ratio measurements are at least as good as  $52$. Similar observations and conclusions with \ac{DL} \ac{RB} utilization results can also be deducted from \ac{FU} ratio observations of Fig.~\ref{FU_Distribution}.

Fig.~\ref{live1} and Fig.~\ref{satellite_1} (second rows marked with yellow colors) depict the \ac{FU} ratios for both terrestrial and satellite eNodeBs respectively.  Note that real \glspl{UE} are connected to the terrestrial eNodeB during the observation period. The scheduling of \glspl{UE} at each \ac{TTI} (i.e. every $1$ ms) may allow some \glspl{UE} to have no resource allocations during this allocation interval. However as depicted in Fig.~\ref{satellite_1}, interestingly the same trend of  \ac{DL} \ac{FU} percentage fluctuations are observed with satellite backhaul network as well. Note that in satellite eNodeB, \ac{FU} percentage values are not expected to fall to zero values since only one test \ac{UE} exists in the satellite network and additionally radio conditions are observed to be good (from observations of \ac{CQI} and \ac{MCS} index values as discussed previously) in satellite eNodeB.

For better visibility of the values in Fig.~\ref{live1} and Fig.~\ref{satellite_1}, we also plot the boxplot for the \ac{FU} ratio (\%) in  \ac{DL} for both terrestrial and satellite eNodeBs  in Fig.~\ref{fu_rb}. In this figure, median values of \ac{FU} ratio for satellite and terrestrial eNodeBs are $60.66 \%$ and $66.87 \%$ respectively. Fig.~\ref{fu_rb} clearly validates a higher \ac{FU} ratio in terrestrial eNodeB compared to satellite eNodeB. Similar to Fig.~\ref{RB_Distribution}, satellite eNodeB cannot schedule the \glspl{PDU} appropriately at \ac{MAC} layer due to existence of incomplete transmit buffer. This in turn reduces \ac{FU} ratios as well as \ac{DL} \ac{PDCP} throughput even though only one test \ac{UE} is connected to satellite eNodeB. An unexpected high variance on \ac{FU} ratio for satellite eNodeB exists  due to existence of unfilled transmit buffer and irregular \ac{PDU} receptions caused by high delay and jitter in satellite link.


\begin{figure} [t]
\centering
\begin{subfigure}{0.5\textwidth}
  \centering
   \includegraphics[width=\linewidth]{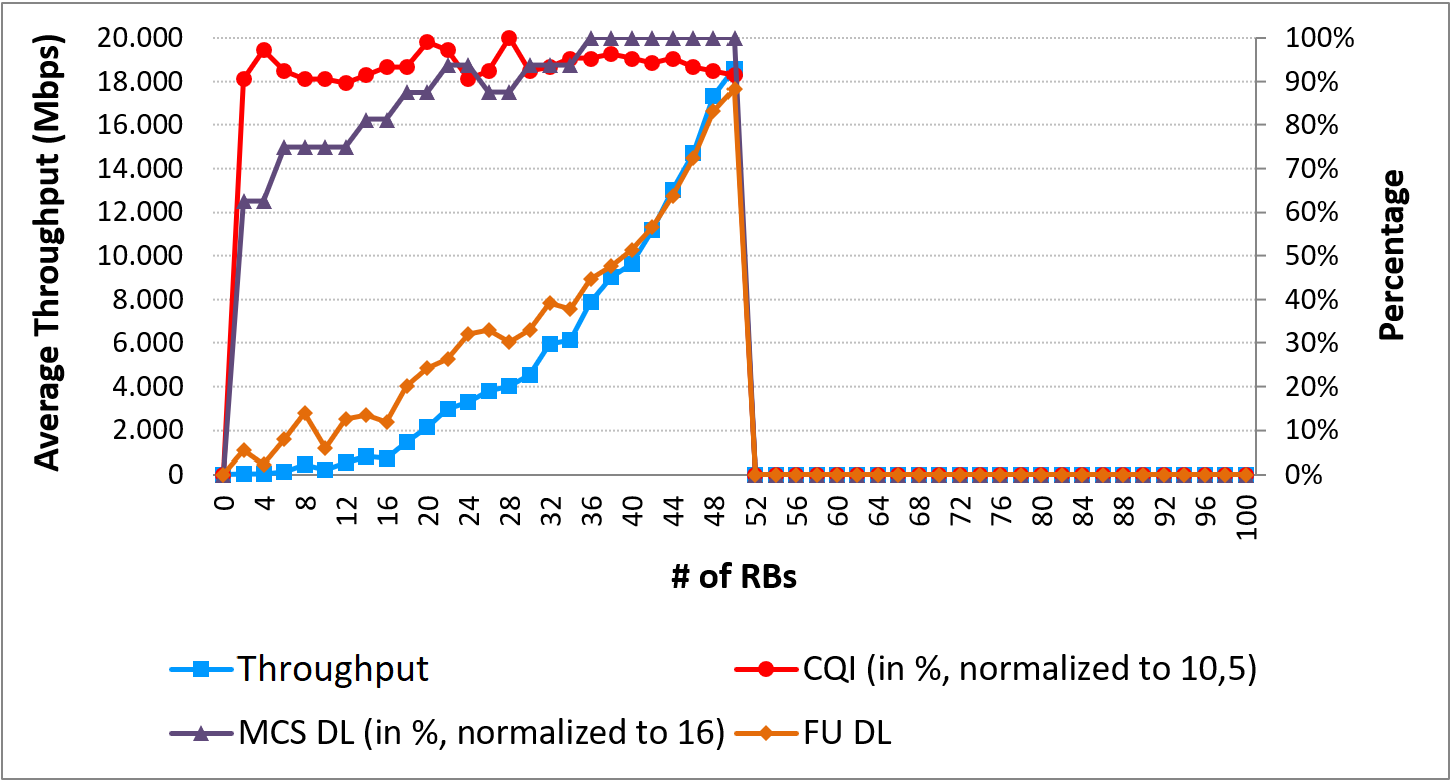}
  \caption{}
  \label{per3}
\end{subfigure} \\
\begin{subfigure}{0.5\textwidth}
  \centering
  \includegraphics[width=\linewidth]{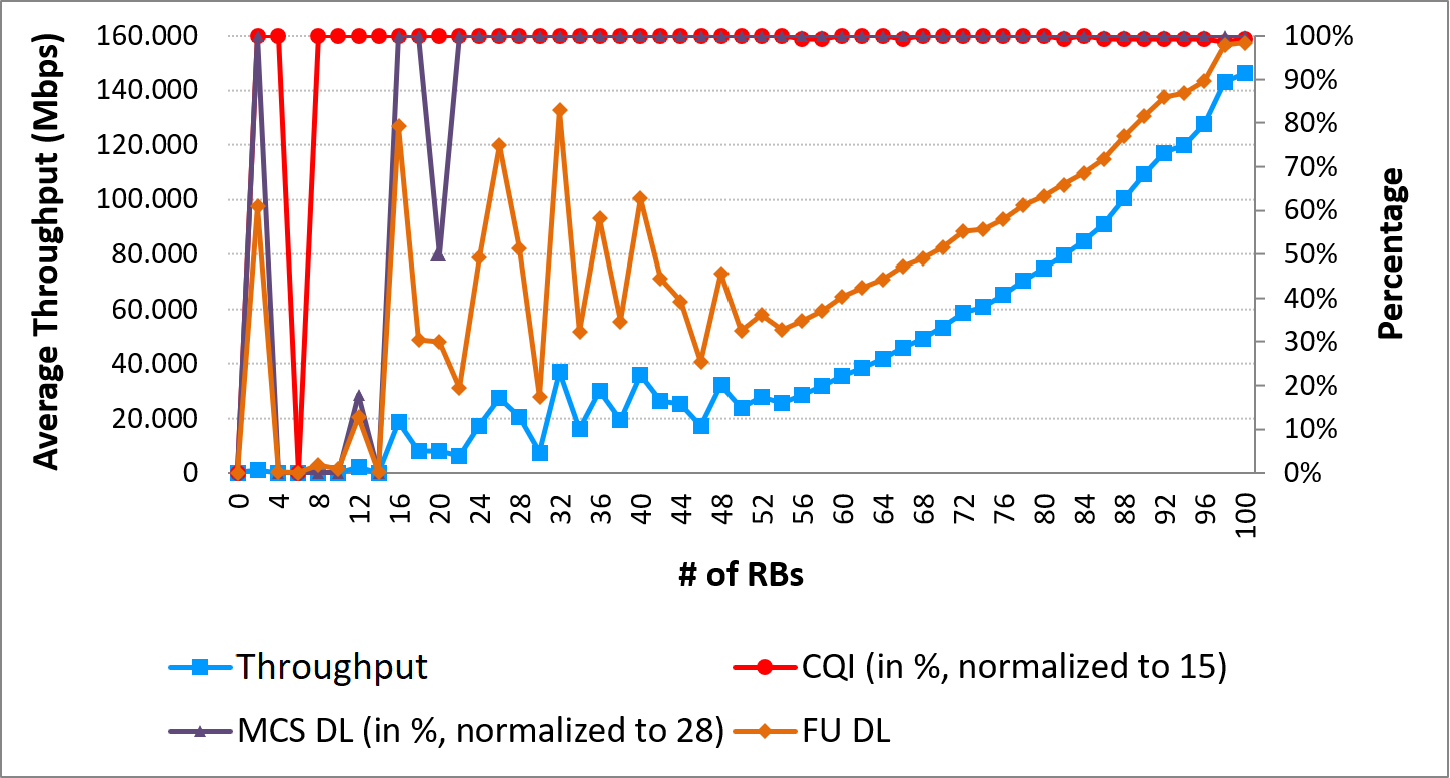}
  \caption{}
  \label{per4}
\end{subfigure}\\
\caption{All KPIs versus \# of RBs utilization relationship analysis for (a) Terrestrial eNodeB (b) Satellite eNodeB [Figures are best viewed in color]}
\label{RB_Relation}
\end{figure}

\textbf{\ac{BLER} percentages:} Light blue marked subfigures in fourth rows of Fig.~\ref{live1} and Fig.~\ref{satellite_1} depict the \ac{DL} \ac{BLER} percentages for terrestrial and satellite eNodeBs respectively. Higher \ac{BLER} percentages are observed in terrestrial eNodeB compared to satellite eNodeB. In general, when \ac{BLER} percentage increases, \ac{MCS} index drops as expected and subsequently throughput decreases. This has been observed in terrestrial eNodeB. Moreover, other \glspl{UE} exist inside terrestrial cell that contribute to higher \ac{BLER} values. When satellite eNodeB results are analyzed, the error detections for \glspl{PDU} that are sent by test \ac{UE}  are recognized very late by \ac{HARQ} due to end-to-end latency in the network. This in turn has increased the \ac{BLER} for satellite eNodeB.

\begin{figure} [t]
\centering
\begin{subfigure}{0.5\textwidth}
  \centering
   \includegraphics[width=\linewidth]{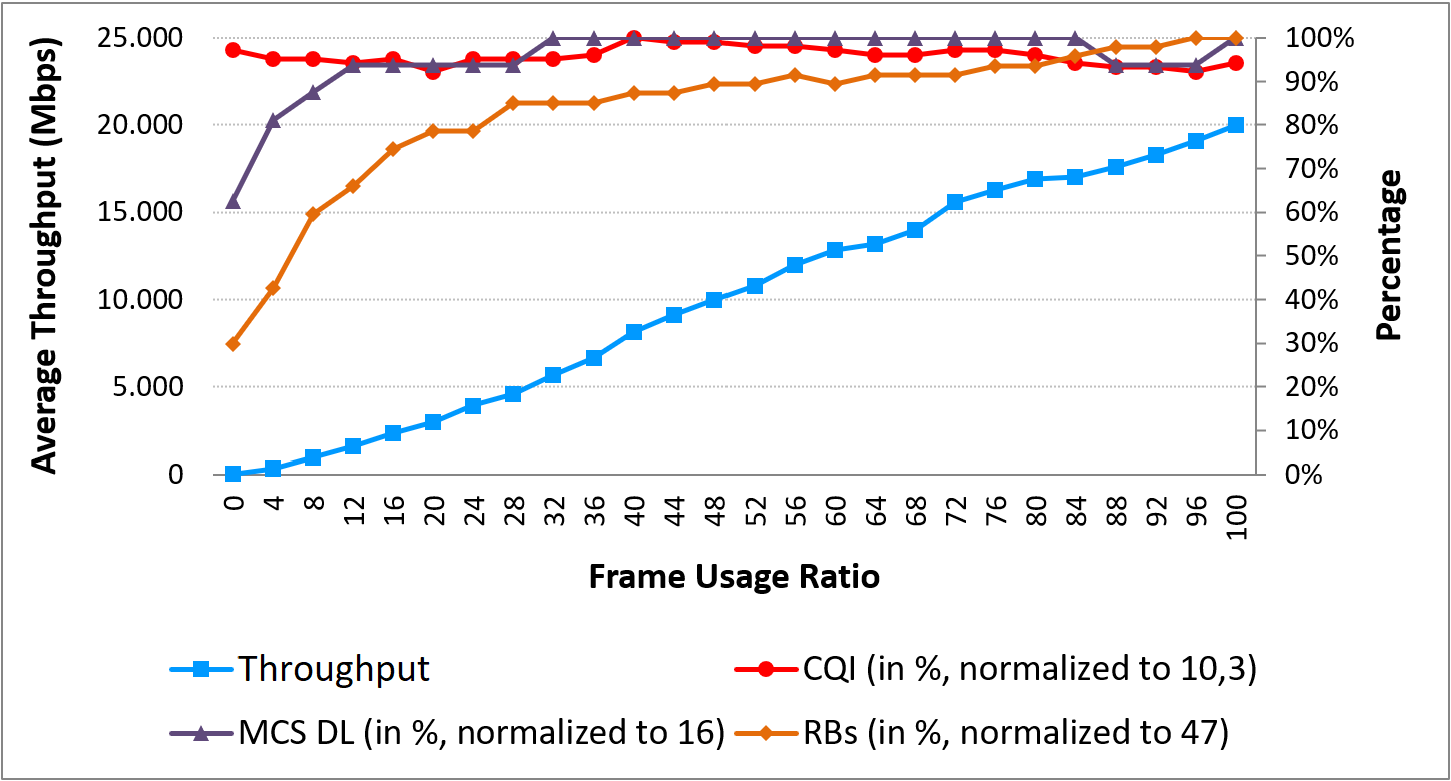}
  \caption{}
  \label{per31}
\end{subfigure} \\
\begin{subfigure}{0.5\textwidth}
  \centering
  \includegraphics[width=\linewidth]{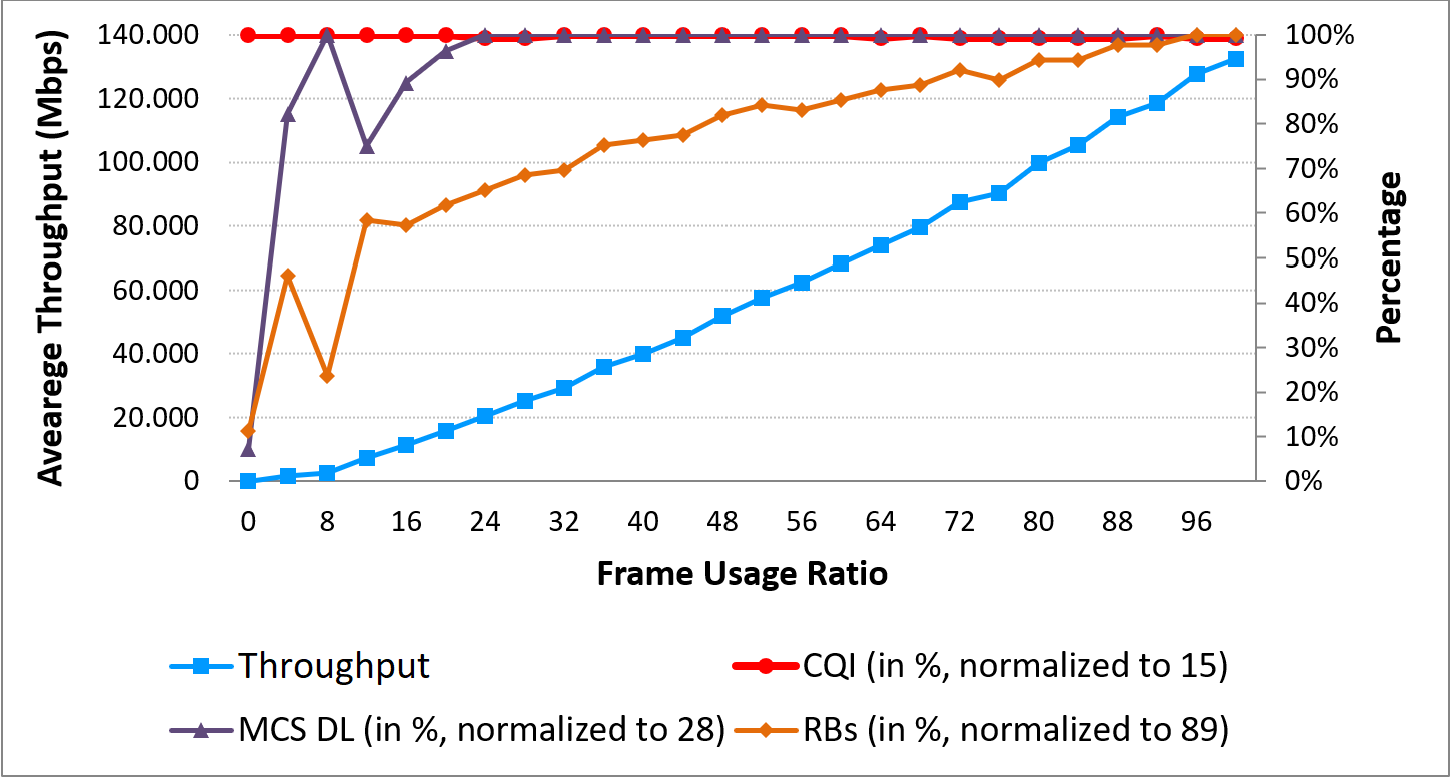}
  \caption{}
  \label{per41}
\end{subfigure}\\
\caption{All KPIs versus DL FU ratio relationship analysis for (a) Terrestrial eNodeB (b) Satellite eNodeB [Figures are best viewed in color]}
\label{FU_Relation}
\end{figure}

\subsection{KPI Relationship analysis}

In this section, we investigate the relationship between different LTE performance \glspl{KPI} by varying one of them and observing its affect on the remaining \glspl{KPI}. Monitored \glspl{KPI} are \ac{DL} throughput, \ac{CQI} values, \ac{FU} ratios, number of \glspl{RB} utilization, \ac{BLER} percentages, number of \glspl{PDU} in \ac{RLC}/\ac{PDCP} layers, \ac{MCS} index values and \ac{MIMO} \ac{TB} usage. In the following analysis, some metrics have been normalized to percentage values so that $100\%$ corresponds to the maximum value. Hence, metrics with different ranges can be represented on the same graph. For instance, \enquote{\glspl{RB} \ac{DL} (in \%, normalized to $30$)} means that the maximum value of \enquote{\glspl{RB} \ac{DL}} in the graph is $30$. In that case, $50\%$ would correspond to \enquote{\glspl{RB} \ac{DL} = $15$}.

\textbf{\ac{KPI} relationship analysis based on number of \glspl{RB} utilization in \ac{DL}:}  Fig.~\ref{RB_Relation} shows the average percentage values for \ac{CQI}, \ac{MCS} index and \ac{FU} ratio on the right y-axis as well as average throughput values on left-y axis versus increasing number of \glspl{RB} utilization on x-axis. Note that in Fig~\ref{per3} and Fig~\ref{per4}, \ac{CQI} values are normalized with respect to maximum value of $10.5$ and $15$ and \ac{DL}  \ac{MCS} index values are normalized with respect to maximum value of $16$ and $28$ respectively. Number of \ac{DL} \glspl{RB} is related to load in the cell. As number of \glspl{RB} increases, throughput values of both satellite and terrestrial eNodeBs also increase accordingly as expected. For example in Fig.~\ref{per3} together with $50$  \glspl{RB} (for $10$ Mhz of bandwidth) around $18$ Mbps can be achieved, whereas in Fig.~\ref{per4} the throughput raises above $140$ Mbps with $100$ \ac{RB} utilization (for $20$ Mhz of bandwidth).

Average \ac{MCS} index and \ac{CQI}  values are at relatively maximum levels for satellite eNodeB in Fig.~\ref{per4}. (Note that high fluctuations exist in Fig.~\ref{per4} at the beginning for average \ac{MCS} index values due to small number of averaged sample data at low number of \glspl{RB}.) Due to \ac{RF} conditions of terrestrial eNodeB, \ac{CQI} and \ac{MCS} index values fluctuate  between  $60\%$ to $100\%$ in Fig.~\ref{per3}.  Comparing \ac{DL} \ac{FU} ratio distributions when the number of \glspl{RB} utilization increases in Fig.~\ref{RB_Relation}, we can observe that the increasing trend is slower in satellite eNodeB compared to terrestrial eNodeB. This is related to transmit buffering problem encountered in satellite eNodeB which is also discussed in more detail in subsequent sections.

\begin{figure*} [ht!]
\centering
\begin{subfigure}{0.8\textwidth}
  \centering
   \includegraphics[width=\linewidth]{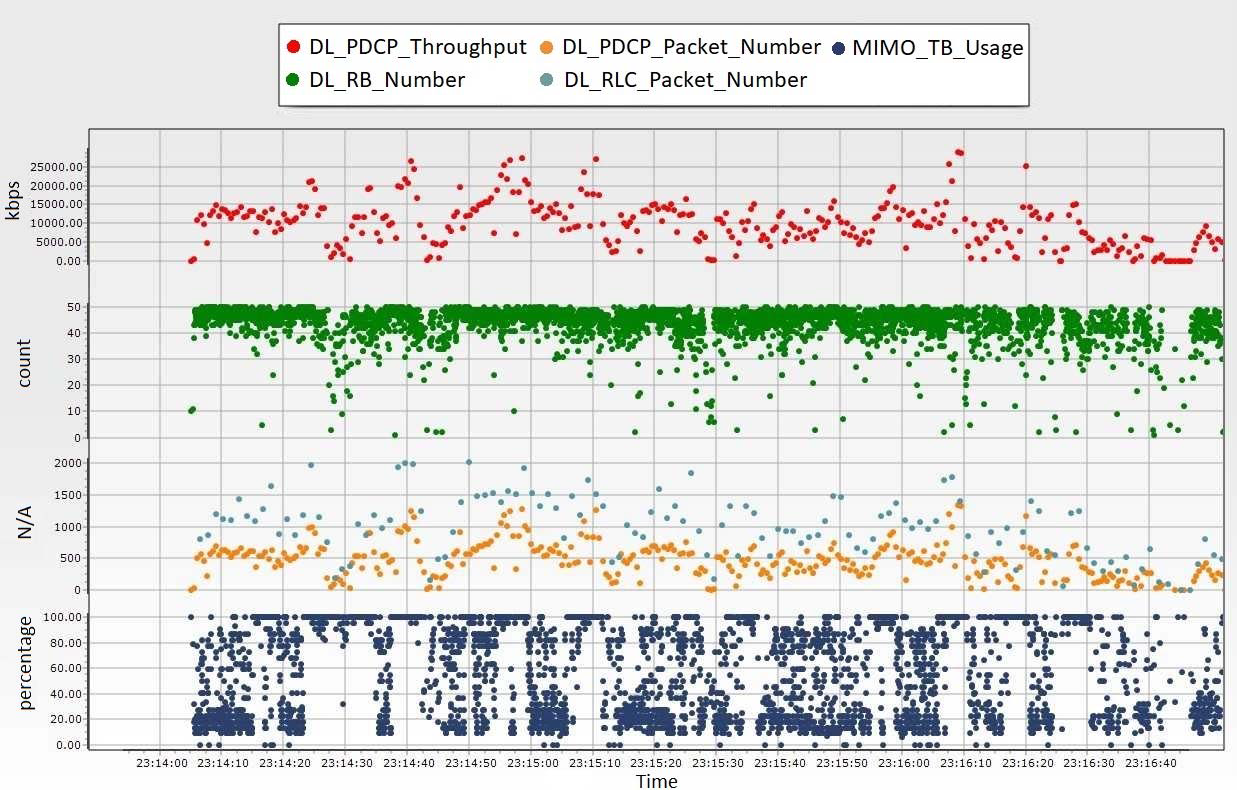}
  \caption{}
  \label{live_2}
\end{subfigure} \\
\begin{subfigure}{0.8\textwidth}
  \centering
  \includegraphics[width=\linewidth]{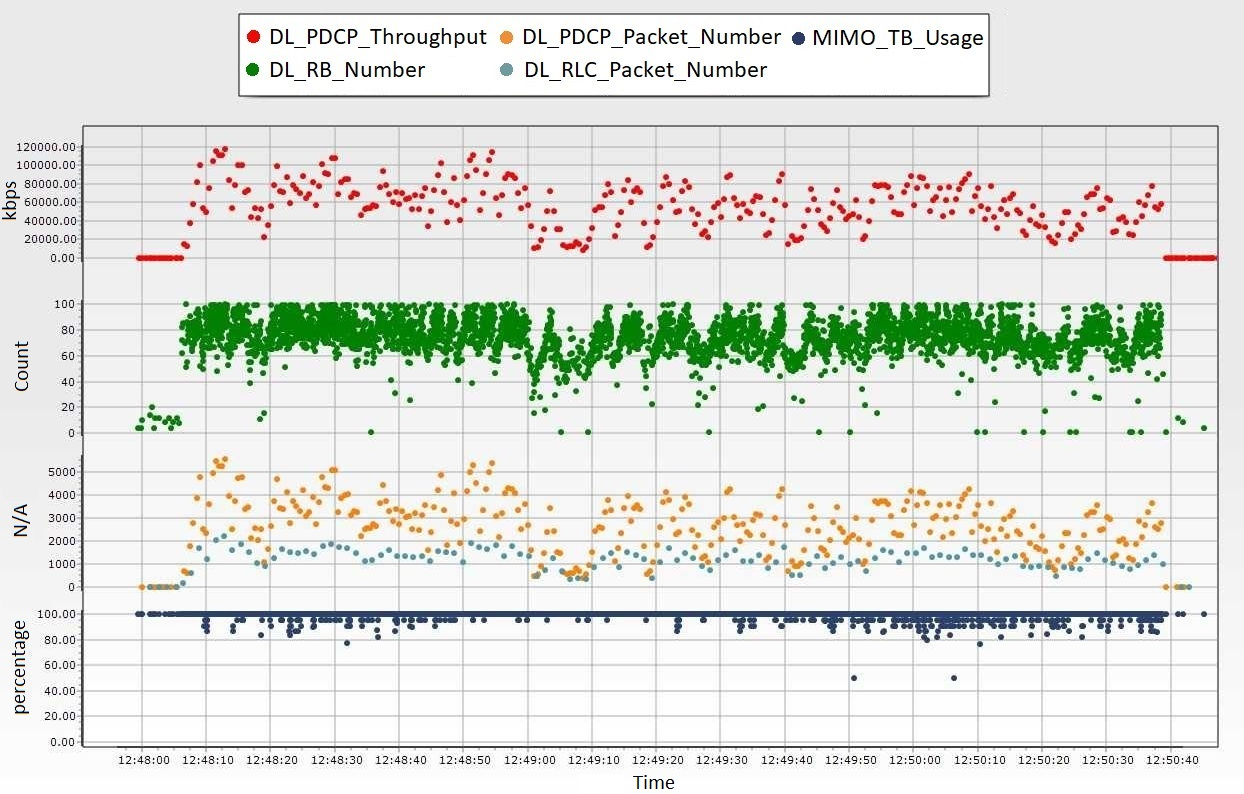}
  \caption{}
  \label{satellite_2}
\end{subfigure}\\
\caption{Additional Radio Performance KPIs in DL for (a) Terrestrial eNodeB (b) Satellite eNodeB [Figures are best viewed in color.]}
\label{tems_kpi_2}
\end{figure*}

\begin{figure} [h]
\centering
   \includegraphics[width=\linewidth]{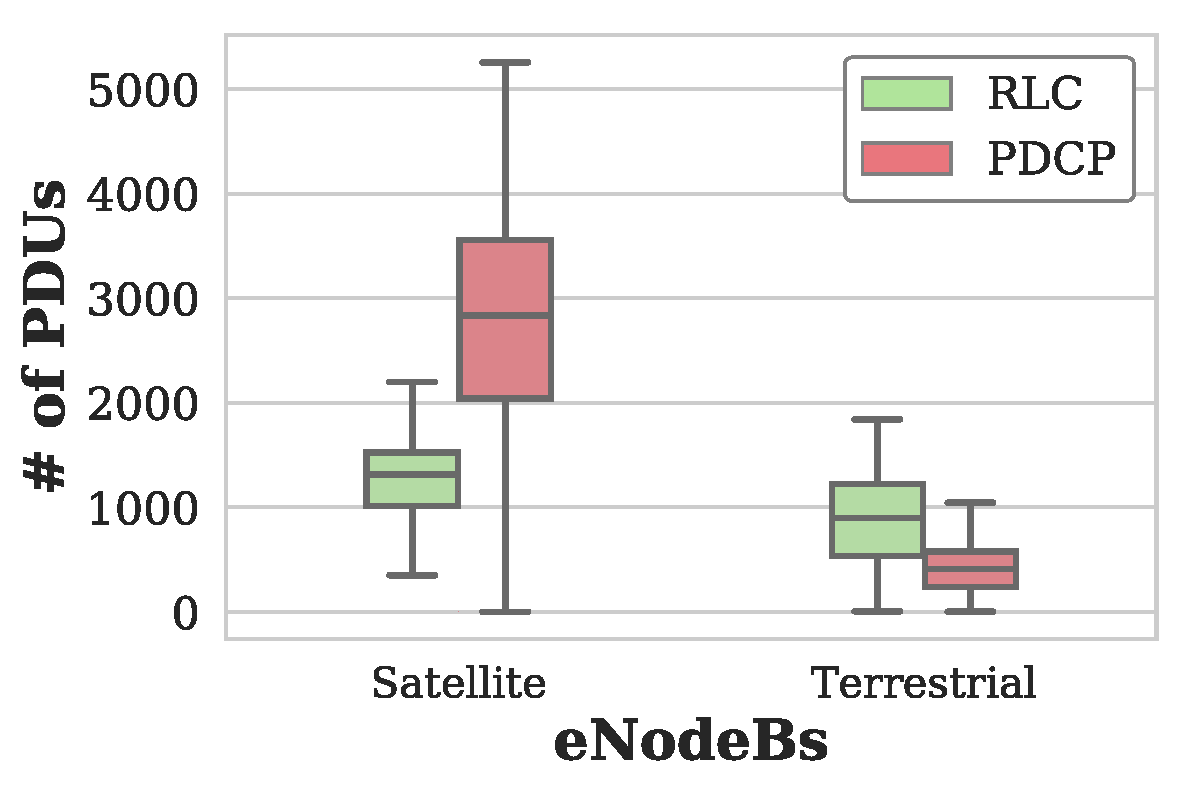}
\caption{Boxplot comparisons of number of PDUs at each radio protocol stack (RLC and PDCP) for both terrestrial and satellite eNodeBs.}
\label{rlc_pdcp}
\end{figure}

\textbf{\ac{KPI} relationship analysis based on \ac{LTE} \ac{DL} \ac{FU} ratio:}  Fig.~\ref{FU_Relation} shows the relationship between \ac{FU} ratio and other related \glspl{KPI}. When  the number of \ac{RB} utilization percentages at a given \ac{FU} ratio (in percentages) are compared between terrestrial and satellite eNodeBs in Fig.~\ref{FU_Relation},  a higher \ac{RB} utilization percentage in terrestrial eNodeB is observed. For example at $32\%$ \ac{FU} ratio, \ac{RB} utilization percentage is around $70\%$ for satellite eNodeB, whereas it reaches to $90\%$ for terrestrial eNodeB.  Fig.~\ref{per41} shows that the number of utilized \glspl{RB}  can decrease as \ac{FU} ratio increases. For instance, the number of \glspl{RB} usage percentage is lower when \ac{FU} ratio is $74$ compared to the case when it is $72$. This result is in contrast with expectations and the observation results of terrestrial eNodeB in Fig.~\ref{per31}. The main reason is again due to the transmit buffer not being filled up completely for appropriate scheduling of \ac{PDCP} \glspl{PDU}.

Fig.~\ref{FU_Relation} also shows the effect of increase in \ac{MCS} and its impact on \ac{PDCP} throughput, \ac{FU} and \ac{CQI}. The effect of jitter and latency in satellite-based backhaul can be extracted from these \glspl{KPI} of satellite eNodeBs. The transmit buffer is not full due to latency originated satellite backhaul link.  For this reason, \ac{PDCP} \glspl{PDU} are kept inside transmit buffer until transmit buffer is full so that they can be scheduled later.  The \ac{PDCP} \glspl{PDU} in the transmit buffer are scheduled to be transmitted quickly due to existence of  high number of \ac{DL} \glspl{RB}. This results in empty \ac{TTI} to be transmitted until new \glspl{PDU} arrive which negatively effects the \ac{FU} ratio (\%) over the satellite link. Due to existence of only one \ac{UE} in satellite eNodeB and the impact of latency and jitter of the satellite link, the \ac{FU} ratio has wide variance as also observed in Fig.~\ref{per41} and Fig.~\ref{fu_rb}. On the other hand, \ac{GTP}-U accelerator yields satisfactory \ac{UE} throughput in contrast to this large variance in \ac{FU} ratio as also observed from Fig.~\ref{per41}. Note that there is only one test \ac{UE} inside satellite network and there is a tendency that \ac{DL} \ac{FU} ratio is expected to worsen in case the number of \glspl{UE} increases.

\textbf{\ac{DL} \ac{PDCP}  throughput:} Fig.~\ref{FU_Relation} marked with purple color shows  \ac{DL} \ac{PDCP}  throughput values. As expected, \ac{DL} \ac{PDCP}  throughput values of both terrestrial and satellite eNodeBs increase as \ac{FU} ratio increases in Fig.~\ref{FU_Relation}. At the same time, satellite eNodeB yields higher average \ac{DL} \ac{PDCP}  throughput values compared to terrestrial eNodeB due to existence of  single \ac{UE} as shown in  last row sub-figures marked with black colors of Fig.~\ref{tems_kpi_1}.

\textbf{\ac{MIMO}-\ac{TB} usage percentage:}  Fig.~\ref{live_2} and Fig.~\ref{satellite_2} show \ac{MIMO}-\ac{TB} usage percentage values which are marked in black color. In both figures, \ac{MIMO}-\ac{TB} values hit $100\%$ which indicate that both terrestrial and satellite eNodeBs utilize \ac{MIMO}. Due to existence of high number of \glspl{UE} in terrestrial eNodeB, \ac{TB} usage percentages have fluctuated over time. On the other hand, no significant \ac{TB} usage percentage change occurs in satellite eNodeB, hence it is around $100\%$.

\textbf{Number of \glspl{PDU} in \ac{PDCP} and \ac{RLC} layers:}   Light blue and yellow colors in third rows of  Fig.~\ref{live_2} and Fig.~\ref{satellite_2} mark the number of \glspl{PDU} in \ac{RLC} and \ac{PDCP} layers respectively. For more detailed analysis, Fig.~\ref{rlc_pdcp} shows the boxplot of the number of \glspl{PDU} in \ac{RLC} and \ac{PDCP} layers for both terrestrial and satellite eNodeBs  over the same observation duration.  We can observe that the median numbers of \ac{DL} \ac{RLC} \glspl{PDU} are  $897$ for terrestrial eNodeB and $1314.5$ for satellite eNodeB.  On the other hand, median numbers of \ac{DL} \ac{PDCP} \glspl{PDU} (that are related to \ac{PDCP} throughput values) are $408.75$ for terrestrial eNodeB and $2836.5$ for satellite eNodeB. This is consistent with \ac{PDCP} throughput values.  The main observation that can be extracted from Fig.~\ref{rlc_pdcp} is that the number \glspl{PDU} in \ac{PDCP} layer is higher than \ac{RLC} layer in satellite eNodeB in comparison with terrestrial eNodeB. This is due to larger waiting period during transmit buffering stage in satellite eNodeB compared to terrestrial eNodeB. In satellite eNodeB, \ac{RLC} layer aggregates more than one \ac{PDCP} packets and sends them to lower layers as one \ac{RLC} packet.

\textbf{Main takeaways and observations:}  Our experimental results reveal three important observations in both  terrestrial and satellite eNodeBs experiments using the \ac{KPI} analysis.

The first observation states that \ac{FU} ratio and \ac{RB} utilization problems caused by the delay and jitter in the satellite link exist.  The variances of \ac{FU} ratio and number of \glspl{RB} distributions are unexpectedly high in satellite eNodeB even though there exists a single test \ac{UE} with high \ac{CQI} and \ac{MCS} index values under satellite eNodeB. The number of \ac{RB} utilization is expected to concentrate around high values in satellite eNodeB. However, \ac{DL} \ac{PDCP} throughput fluctuates in satellite eNodeB as better evidenced in Fig.~\ref{satellite_2}. The main reason for this phenomenon is that arriving \ac{GTP} \glspl{PDU} over satellite links are not regularly filling out the transmit buffer at satellite eNodeB. If the transmit buffer size becomes full, satellite eNodeB will be able to schedule and send \ac{PDCP} \glspl{PDU} appropriately to \glspl{UE} in \ac{DL}. However, this is not observed most of the time during our observation period. For this reason, distribution of number of \ac{RB} utilization is more dispersed than expected. These unbalanced distributions in both \ac{FU} ratio and number of \ac{RB} utilization are mainly due to delay and jitter of satellite link.

The second observation states that in satellite link relatively low numbers of \glspl{PDU} are generated at the \ac{RLC} layer compared to the \ac{PDCP} layer. The \ac{CQI} values of the test \ac{UE} for satellite are high, however high numbers of \ac{PDCP} \glspl{PDU} are also generated. The scheduler at the satellite eNodeB is concatenating \ac{PDCP} \glspl{PDU} and sending them to  \ac{MAC} layer as one \ac{SDU}. However, this process of scheduler is delayed due to effect of satellite link. Under normal circumstances, schedulers are expected to segment \ac{PDCP} \glspl{PDU} at the \ac{RLC} layer. However, this kind of segmentation does not appear in satellite eNodeB where \ac{RLC} \glspl{PDU} are segmented at the \ac{PDCP} layer. For this reason, in case many \glspl{UE} inside satellite eNodeB exist, the existence of high number of \ac{PDCP} \glspl{PDU} can create a transmit buffering problem.

The third observation states that the excessive existence of \ac{PDCP} \glspl{PDU} can be due to \ac{GTP}-U accelerator. \ac{GTP}-U accelerator technique includes a caching method. Therefore, we observe that there can be a trade-off between the number of \glspl{UE} that can be supported within a eNodeB with satellite-based backhaul link and the caching size related to the \ac{GTP}-U accelerator's performance. Large caching size increases the capabilities of the \ac{GTP}-U accelerator in terms of providing lower latency. On the other hand, it can also decrease the number of \glspl{UE} that can be scheduled over the satellite eNodeB due to existence of large number of \ac{PDCP} \glspl{PDU} compared to \ac{RLC} \glspl{PDU}. Lower caching size on the other hand adversely effects the \ac{GTP}-U's accelerator, but decreases the number of \ac{PDCP} \glspl{PDU} in the transmit buffer of the scheduler. Therefore, for higher performance gains in terms of low latency and high support of \glspl{UE} for satellite eNodeB, an optimal balance between the caching size and the number of \glspl{UE} using satellite eNodeB is needed. The reason is that \ac{GTP}-U acceleration is a fundamental requirement for satellite links to decrease the inherent latency.

\section{Conclusions}
\label{conclusions}

It is inevitable that satellite networks will integrate with other networks including \ac{5G} cellular networks. In this paper, we investigated various cellular network \glspl{KPI}  of both satellite and terrestrial eNodeBs in an experimental set-up to observe the effect of satellite-based backhaul links on radio network performance.  The comparative experimental performance evaluation is  performed in terms of \ac{CQI} values, \ac{MCS} index, \ac{PDCP} throughput, number of utilized \glspl{RB}, \ac{FU} utilization ratios as well as \ac{BLER} and \ac{MIMO} \ac{TB} utilization.  Our experimental results indicate three main observations: The first one is the existence of \ac{RB} and \ac{FU} utilization problems in satellite-based backhaul network due to inherent delay and jitter on satellite links even though high \ac{MCS} index and \ac{CQI} values are monitored during observation intervals. The second one is that compared to terrestrial network total number of \ac{PDCP} packets outnumbers the total number of \ac{RLC} packets in satellite-based backhaul network. Our final observation designates that there will be a trade-off between the number of \glspl{UE} that can be supported with satellite eNodeB and the caching size of the utilized \ac{GTP}-U accelerator. Therefore, an optimal balance between the caching size of the \ac{GTP}-U accelerator for latency reductions and the number of \glspl{UE} using satellite eNodeB needs to be adjusted before deployment of satellite-based backhaul networks in an operational environment.

\ifCLASSOPTIONcaptionsoff
  \newpage
\fi

\section*{Acknowledgments}
This work was partially funded by Spanish MINECO grant TEC2017-88373-R (5G-REFINE) and by Generalitat de Catalunya grant 2017 SGR 1195.

\bibliographystyle{unsrt}
\bibliographystyle{abbrv}
\bibliography{References}

\end{document}